\documentclass[preprint,showpacs,showkeys,aps,prd,amsfonts,eqsecnum,nofootinbib]
{revtex4} 
\usepackage{graphicx,epsfig}
\newcommand{\Rslash}{R\kern -6.4pt\big{/}}

\begin{document}
\preprint{CUMQ/HEP 137}
%
%
\title{CP-ODD  PHASE EFFECTS IN A LEFT-RIGHT SYMMETRIC CHARGINO SECTOR}
\author{Mariana Frank}
\email[]{mfrank@vax2.concordia.ca}
\author{Ismail Turan}
\email[]{ituran@physics.concordia.ca}
\author{Artorix de la Cruz de O\~{n}a}
\email[]{artore-d@yahoo.com}
\affiliation{Department of Physics, Concordia University, 7141 Sherbrooke West Street, Montreal, Quebec, CANADA H4B 1R6}

\date{\today}

\begin{abstract}
The left-right supersymmetric model contains a right-handed gaugino, as well as several higgsinos, in addition to the minimal supersymmetric model. Thus several CP-noninvariant phases appear in this sector. We analyze their impact on chargino masses and find that only two combinations are physically relevant. We then study the production of charginos in $e^+e^-$ annihilation and chargino decays into a sneutrino and a lepton, and investigate the effects of CP-phases. We also study the CP-odd asymmetry in the production and subsequent decay at the linear collider with longitudinally polarized beams and find a large enhancement when the decay channel to the right sneutrino is available. The effects of the phases  in the left-right supersymmetric  chargino sector are different from the minimal supersymmetric standard model, and signals from this sector would be able to distinguish between different gauge symmetries.
\end{abstract}
\vskip -2.5cm
\maketitle
\section{Introduction}\label{sec:intro}

Charge-parity (CP) violation is one of the least understood phenomena in high energy physics. 
In the standard model (SM), CP violation is parametrized by one arbitrary phase in the quark mixing matrix 
and its value fixed by experimental observations. Models beyond the SM, especially those 
including supersymmetry, predict several new CP-odd phases. The effects of some of these phases, 
particularly the ones coming from the soft supersymmetry breaking scalar masses, have been 
studied extensively \cite{Ellis:1982tk}. In the minimal supersymmetric standard model (MSSM), these phases 
need to either be unreasonably small, or the scalar fermion masses unusually large \cite{kizukuri}.

A solution to this so called ``SUSY  CP-problem'' has been suggested by the authors 
in \cite{Mohapatra:1995xd}. By enlarging the supersymmetric gauge sector to be left-right symmetric, parity forces Yukawa couplings to be Hermitean, and mass terms in the superpotential to be real. Thus the CP problem is solved at the right-handed ($SU(2)_R$, or seesaw) scale $M_{\Delta_R}$. If $M_{\Delta_R}$ is of the order of the electroweak scale, then no other phase can be generated and the problem is solved. If however $M_{\Delta_R}$ is a higher scale, the scenario favored by neutrino masses and grand-unification schemes with left-right models as intermediate steps, this is no longer the case. In some variants of the model, the phases appear only at two loop level and are thus naturally small \cite{Mohapatra:1995xd}, in others the electroweak phase is generated form the renormalization group equations evolution from the seesaw scale, and the electron dipole moment (EDM) problems persist \cite{Frank:1999sy}.

The left-right supersymmetric model (LRSUSY) enlarges the minimal supersymmetric standard model (MSSM) gauge symmetry  $SU(3)_C \times SU(2)_L \times U(1)_Y$ to   $SU(3)_C \times SU(2)_L \times SU(2)_R \times U(1)_{B-L}$. This symmetry allows for the seesaw 
mechanism within supersymmetry and predicts neutrino masses and 
mixing naturally. In the supersymmetric sector of the theory, it introduces right-handed gauginos in addition to several higgsinos. We expect this change to be most pronounced in the electroweak chargino sector, where previously forbidden right-right interactions are now allowed to proceed with a strength comparable to left-left interactions. This could have significant effects on the CP-phases in the sector, where more phases are allowed, which could distinguish the model from the MSSM. By contrast, we expect the effect to be less pronounced in the neutralino sector, where right-handed gaugino interactions are permitted in MSSM.

 In this work, we  investigate the effects of such phases on the masses, production, and decay rates of charginos. Our emphasis is on an analysis of a general unconstrained model, assuming arbitrary phases, and finding the combination of angles that would affect the phenomenology, as well as presenting  experimentally testable signals from the chargino sector. We assume in what follows that the phases can be taken arbitrarily large, and that CP-violation restrictions in the quark and lepton sectors can be satisfied either by allowing scalar masses to be large, or through cancellations of various contributions \cite{Ibrahim:1999af}.
 
 We perform a detailed and complete analysis of the chargino sector including masses, production rates, forward-backward asymmetry in production and decays. For decays, we chose to analyze the channel most sensitive to left-right symmetry, the decay into a sneutrino and a lepton. We analyze the dependence on production and subsequent decay from a polarized $e^+e^-$ beam and the T-odd asymmetry associated with this.
  
Our paper is organized as follows: In Section \ref{sec:model} we present 
the general framework of the left-right supersymmetric model with special emphasis on the chargino sector. 
We then  
discuss the chargino masses, with particular attention paid to the effect of the CP-odd angles  in Section \ref{sec:masses}. In Section \ref{sec:production}, we 
investigate the cross section for chargino production in an unpolarized $e^+e^-$ collider and its dependence on the phases in the theory.
We discuss the variation of the decay rates of charginos with CP-phases as well as the T-odd asymmetry in polarized production and decay in Section \ref{sec:decays}. Section \ref{sec:conclusion} is devoted to our summary and conclusions.

\section{\bf The chargino sector of the LRSUSY Model}\label{sec:model}

The minimal supersymmetric left-right model is based on the gauge group
$SU(3)_C \times SU(2)_L \times SU(2)_R \times U(1)_{B-L}$ \cite{Francis:1990pi}. The matter
fields of this model consist of three families of both left and right handed quark $Q_L, Q_R$ and lepton 
$L_L, L_R$ fields. The Higgs
sector consists of two bidoublet ($\Phi_u,~\Phi_d$) and four triplet Higgs ($\Delta_L, ~\Delta _R$ and $\delta_L, ~\delta _R$) superfields. 
The bidoublet  Higgs superfields 
implement the $SU(2)_L \times U(1)_{Y}$ symmetry breaking and generate
a Cabibbo-Kobayashi-Maskawa (CKM) mixing matrix. The triplet Higgs
$\Delta_L, ~\Delta_R$ bosons allow for the seesaw
mechanism, while the additional
triplet fields $\delta_L, ~\delta_R$ are needed to cancel triangle
gauge anomalies in the fermionic sector. The most general superpotential
involving these superfields is:
\begin{eqnarray}
\label{superpotential}
W & = & {\bf Y}_{Q}^{(i)} Q_L^T\Phi_{i}i \tau_{2}Q_R + {\bf Y}_{L}^{(i)}
L_L^T \Phi_{i}i \tau_{2}L_R + {\bf Y}_{LR}(L_L^T i \tau_{2} \delta_L L_L +
L_R^{T} i \tau_{2}
\Delta_R L_R) \nonumber \\
& & + \mu_{LR}\, {\rm tr}(\Delta_L  \delta_L +\Delta_R
\delta_R) + \mu_{ij}\,{\rm tr}(i\tau_{2}\Phi^{T}_{i} i\tau_{2} \Phi_{j})
+W_{NR},
\end{eqnarray}
where ${\bf Y}_Q$ and  ${\bf Y}_L$ are
the Yukawa couplings
for the quarks  and leptons, respectively and ${\bf Y}_{LR}$ is the coupling for the triplet Higgs bosons.  The parameters $\mu_{ij}$ and $\mu_{LR}$ are the Higgs mass parameters.\footnote{ Strictly speaking, the left-right symmetry is defined up to a rotation in family space and quark and lepton labels are meaningful only after defining Yukawa couplings and Higgs vevs.} Left right symmetry requires all ${\bf Y}$-matrices to be Hermitean in the generation space  and  ${\bf Y}_{LR}$ matrix to be symmetric. 
Here $W_{NR}$ denotes (possible) non-renormalizable terms arising 
from higher scale
physics or Planck scale effects \cite{Chacko:1997cm}. The presence of these terms
insures that, when the SUSY breaking scale is above $M_{W_{R}}$, the
ground state is R-parity conserving. Left-right supersymmetric models have one additional problem: two of the physical Higgs bosons can contribute flavor-violating neutral currents (FCNC) by violating flavor by two units and by generating too large a contribution to the $K_L-K_S$ mass splitting. This problem can be circumvented by raising the right-handed mass scale to $M_{W_R}>2.5-5$ TeV \cite{Pospelov:1996fq}. In addition, the
Lagrangian also includes soft supersymmetry breaking terms as well as  $F$- and, $D$-terms. 
For a more detailed description of the model, see \cite{Frank:2005vd}.
 
In the supersymmetric sector of the model there are six 
singly-charged
charginos, corresponding to $\tilde\lambda_{L}$,
$\tilde\lambda_{R}$, $\tilde\phi_{u}$,
$\tilde\phi_{d}$, $\tilde\Delta_{L}^{\pm}$, and
$\tilde\Delta_{R}^{\pm}$.
The model also has eleven neutralinos, corresponding to
$\tilde\lambda_{Z}$,
$\tilde\lambda_{Z^{\prime}}$,
$\tilde\lambda_{V}$,  $\tilde\phi_{1u}^0$, $\tilde\phi_{2u}^0$,
$\tilde\phi_{1d}^0$,  $\tilde\phi_{2d}^0$, $\tilde\Delta_{L}^0$,
$\tilde\Delta_{R}^0$,  $\tilde\delta_{L}^0$, and
$\tilde\delta_{R}^0$.  The doubly charged Higgs
and Higgsinos do not
affect quark phenomenology, but the neutral and singly charged
components do, through
mixings in the chargino and neutralino mass matrices.\footnote {Note however that the doubly charged Higgs and higgsinos contribute to lepton phenomenology \cite{Frank:2000dw}.}
The terms relevant to the masses of charginos in the Lagrangian are
\begin{equation}
{\cal L}_C=-\frac{1}{2}(\psi^{+T}, \psi^{-T}) \left ( \begin{array}{cc}
                                                        0 & X^T \\
                                                        X & 0
                                                      \end{array}
                                              \right ) \left (
\begin{array}{c}
                                                               \psi^+ \\
                                                               \psi^-
                                                               \end{array}
                                                        \right ) + {\rm H.c.} \ ,
\end{equation}
where $\psi^{+T}=(-i \lambda^+_L, -i \lambda^+_R, \tilde{\phi}_{1d}^+,
\tilde{\phi}_{1u}^+, \tilde{\Delta}_R^+)$
and $\psi^{-T}=(-i \lambda^-_L, -i \lambda^-_R, \tilde{\phi}_{2d}^-,
\tilde{\phi}_{2u}^-, \tilde{\delta}_R^-)$, and
\begin{equation}
X=\left( \begin{array}{ccccc}
                            M_L & 0 & 0 & g_L\kappa_d & 0
\\
           0 & M_R &0  & g_R\kappa_d 
&\sqrt{2}g_Rv_{\delta_R}
\\
    g_L\kappa_u & g_R \kappa_u & 0 &-\mu &  0
\\
0 & 0  & -\mu &0 & 0\\
       0 & \sqrt{2} g_R v_{\Delta_R} & 0 & 0 & -\mu
               \end{array}
         \right ),
\end{equation}
where we have taken, for simplification, $\mu_{ij}=\mu$. Here $\kappa_u$ and $\kappa_d$ are the bidoublet Higgs bosons vacuum expectation values (VEVs), $v_{\Delta_R}$ and $v_{\delta_R}$ are the triplet Higgs bosons VEVs, and $M_{L}, M_{R}$ the $SU(2)_L$  and $SU(2)_R$ gaugino masses, respectively. The chargino mass
eigenstates $\chi_i$ are obtained by
\begin{eqnarray}
\chi_i^+=V_{ij}\psi_j^+, \ \chi_i^-=U_{ij}\psi_j^-, \ i,j=1, \ldots 5,
\end{eqnarray}
with $V$ and $U$ unitary matrices satisfying
\begin{equation}
U^* X V^{-1} = M_D.
\end{equation}
The diagonalizing matrices $U^*$ and $V$ are obtained by
computing the eigenvectors corresponding
to the eigenvalues of $X X^{\dagger}$ and $X^{\dagger} X$, respectively.

\section{\bf Chargino masses with CP-odd  phases in the LRSUSY }\label{sec:masses}

From the above general mass matrices for charginos, we can obtain analytic expressions for the eigenvalues. First, note  that the VEVs of the triplet Higgs bosons, $v_{\Delta_R}$ and $v_{\delta_R}$, are responsible for giving masses to $W_R, ~Z_R$ bosons, as well as implementing the seesaw mechanism. As such, $ v_{\Delta_R} \gg $ several TeV, and thus we can safely decouple $\Delta_R^+$ and $\delta_R^-$ from the lighter chargino states.

We now incorporate the most general set of CP violating phases in the mass matrix. That is, we do not restrict the phases according to any particular scenario, and allow the most general set of phases. These can easily be generated from the RGE  evolution running down from $M_{\Delta_R}$ to the electroweak scale. Since $M_{L}, M_{R}$  and $\mu$, the Higgsino coupling parameter are assumed complex at the electroweak scale, with nontrivial phases $\xi_{1},\xi_{2}$ and $\theta_{\mu}$, respectively, then they can include  phases: 
\begin{eqnarray}
M_{L}\equiv |M_{L}| e^{i\xi_{1}},\qquad M_{R}\equiv|M_{R}|e^{i\xi_{2}},\qquad \mu\equiv|\mu |e^{i\theta_{\mu}}, 
\end{eqnarray} 
and, in addition we allow the Higgs vacuum expectation values to be complex, with phases $\chi_1,~\chi_2$: $\langle \phi_u \rangle= |\kappa_u| e^{-i \chi_1}$ and $\langle \phi_d \rangle= |\kappa_d| e^{-i \chi_2}$. The chargino mass matrix $X$ with the most general allowed set of $CP$ violating phases becomes
\begin{eqnarray} 
X=\left(\begin{array}{cccc}|M_{L}|e^{i\xi_{1}} & 0 & 0 & g_{L}\,|\kappa_{d}|e^{-i\chi_{2}}\\0 & |M_{R}|e^{i\xi_{2}} & 0 & g_{R}\,|\kappa_{d}|e^{-i\chi_{2}}\\g_{L}\,|\kappa_{u}|e^{-i\chi_{1}} & g_{R}\,|\kappa_{u}|e^{-i\chi_{1}} & 0 & -|\mu| e^{i\theta_{\mu}}\\0 & 0 & -|\mu| e^{i\theta_{\mu}}& 0\end{array}\right),
\end{eqnarray}
where the phase angles (i.e. $|\cos \theta_{\mu}|\leq 1$) are in the interval $[0,2\pi]$. 
The results in the $CP-$ conserving limit, maybe obtained by simple taking the values $0$ or $\pi$, in the  matrix. Using the transformation
\begin{eqnarray}
X=P^T_{\chi}M^{c}P_{\chi},
\end{eqnarray}
where
\begin{eqnarray} 
P_{\chi}=\left(\begin{array}{cccc}e^{-i\xi_{1}/2} & 0 & 0 & 0\\0 & e^{-i\xi_{2}/2} & 0 & 0\\0&0&e^{i(\xi_{1}/2+\chi_{1})}&0 \\0 & 0 &0& e^{i(\xi_{2}/2+\chi_{2})}
\end{array}\right)
\end{eqnarray}
is a unitary matrix. Then the matrix $M^{c}$ can be written as
\begin{eqnarray} 
M^{c}=\left(\begin{array}{cccc}|M_{L}| & 0 & 0 & g_{L}\,|\kappa_{d}|e^{-i\xi}\\0 & |M_{R}|& 0 & g_{R}\,|\kappa_{d}|\\g_{L}\,|\kappa_{u}|& g_{R}\,|\kappa_{u}|e^{i\xi} & 0 & -|\mu| e^{i\theta}\\0 & 0 & -|\mu| e^{i\theta}& 0\end{array}\right),\label{eq:charmassmatrix}
\end{eqnarray}
where we have defined $\theta=(\xi_{1}+\xi_{2})/2+\theta_{\mu}+\chi_{1}+\chi_{2}$ and $\xi=(\xi_{1}-\xi_{2})/2$. These phases allow for both explicit and/or spontaneous CP violation; in the latter case $\theta=\chi_{1}+\chi_{2}$ and $\xi=0$.\footnote{ In the left-right symmetric model, spontaneous CP violation only has been shown to provide insufficient CP violation in the B decays \cite{Ball:1999mb}. }The dependence with respect to only two angles is now clear. The phases $\theta$ and $ \xi$ are the new sources of $CP$ violation, which can vary in the range $0\leq \theta \leq 2\pi$, and $0\leq \xi \leq \pi$, respectively.
To diagonalize the matrix $M^{c}$ we use the following transformation
\begin{eqnarray}
\label{eq:g1}V\left[(M^{c})^{\dag}M^{c}\right]V^{-1} ={\rm diag}(\mathrm{\tilde{M}}_{\chi^{\pm}_{1}},\mathrm{\tilde{M}}_{\chi^{\pm}_{2}},\mathrm{\tilde{M}}_{\chi^{\pm}_{3}},\mathrm{\tilde{M}}_{\chi^{\pm}_{4}})\equiv M_{D}^{2},
\end{eqnarray}
where we assume $\mathrm{\tilde{M}}_{\chi^{\pm}_{1}}\le\mathrm{\tilde{M}}_{\chi^{\pm}_{2}}\le\mathrm{\tilde{M}}_{\chi^{\pm}_{3}}\le\mathrm{\tilde{M}}_{\chi^{\pm}_{4}}$.
Notice that the transformation matrix $V$ is now a $4 \times 4$ matrix and a function only of $\theta$ and $\xi$. The chargino masses can be determined by solving the eigenvalue equation analytically. We use the method outlined in \cite{Gounaris:2001fx}. We get the exact analytic expressions for the chargino masses as functions of the CP angles, i.e., $ \mathrm{\tilde{M}}_{\chi^{\pm}_{i}} = \mathrm{\tilde{M}}_{\chi^{\pm}_{i}}(\theta, \xi),~i=1, \ldots, 4$. 
 The analytic formulas for the chargino masses are given by
\begin{eqnarray}
\mathrm{\tilde{M}}_{\chi^{\pm}_{1}} &=&{\cal R}e\left[\frac{a}{4}-\frac{\alpha}{2}-\frac{1}{2}\,\sqrt{\zeta-\varpi-\frac{\lambda}{4\alpha}}\right]^{\frac{1}{2}},\\
\mathrm{\tilde{M}}_{\chi^{\pm}_{2}} &=&{\cal R}e\left[\frac{a}{4}-\frac{\alpha}{2}+\frac{1}{2}\,\sqrt{\zeta-\varpi-\frac{\lambda}{4\alpha}} \right]^{\frac{1}{2}},\\
\mathrm{\tilde{M}}_{\chi^{\pm}_{3}} &=&{\cal R}e\left[\frac{a}{4}+\frac{\alpha}{2}-\frac{1}{2}\,\sqrt{\zeta-\varpi+\frac{\lambda}{4\alpha}} \right]^{\frac{1}{2}},\\
\mathrm{\tilde{M}}_{\chi^{\pm}_{4}} &=&{\cal R}e\left[\frac{a}{4}+\frac{\alpha}{2}+\frac{1}{2}\,\sqrt{\zeta-\varpi+\frac{\lambda}{4\alpha}}\right]^{\frac{1}{2}},
\end{eqnarray}
where ${\cal R}e$ represents the real value of the function and 
\begin{eqnarray} 
\alpha&\equiv& \nonumber \sqrt{\beta+\nu+\varpi},\:\quad\beta \equiv\nonumber\left[\frac{a^{2}}{4}-\frac{2b}{3}\right],\\
\zeta &\equiv& \nonumber \left[\frac{a^{2}}{2}-\frac{4b}{3}-\nu\right],\,\,\;\nu \equiv\nonumber\frac{(2^{\frac{1}{3}}\,\gamma)}{3\,\epsilon},\\
\varpi &\equiv& \nonumber\frac{\epsilon}{32^{\frac{1}{3}}},\,\,\,\epsilon \equiv \nonumber(\delta+\sqrt{\eta})^{\frac{1}{3}},\:\quad\qquad\eta \equiv \nonumber (-4\,\gamma^{3}+\delta^{2}),\\
\gamma &\equiv& \nonumber (b^{2}+3\,a\,c+12\,d),\;\delta \equiv\nonumber(2\,b^{3}+9\,a\,b\,c+27\,c^{2}+27\,a^{2}\,d-72\,b\,d),\\
\lambda &\equiv&\nonumber (a^{3}-4\,a\,b-8\,c),
\end{eqnarray}
with parameters $a,~b,~c,~d$ given in terms of the chargino mass-squared matrix elements $M_{ij}=M^{c\,\dagger}_{ik} M^{c} _{kj}$ as :
\begin{eqnarray} 
a  &\equiv& {\rm tr} M_{ij}, ~ b \equiv \frac{1}{2}\left [ ({\rm tr}M_{ij})^2-{\rm tr}M^2_{ij})\right ], \nonumber \\
c   &\equiv& \frac16\left [ ({\rm tr}M_{ij})^3- 3 {\rm tr}M_{ij}{\rm tr}M^2_{ij}+2 {\rm tr}M^3_{ij}\right ],~
d  \equiv{\det} M_{ij}.
\end{eqnarray} 
Similarly, we can obtain analytic expressions for the diagonalizing matrices $V_{ij},~U^\ast_{ij}$. While expressions for these exist in the literature \cite{Gounaris:2001fx, Choi:2001ww}, here we will fix the masses and $\tan \beta$ and use the numerical values for the elements which include the phases.

\begin{figure}[h]
\vspace{-0.05in}  
    \centerline{\hspace*{3.1cm} \epsfxsize 3.5in {\epsfbox{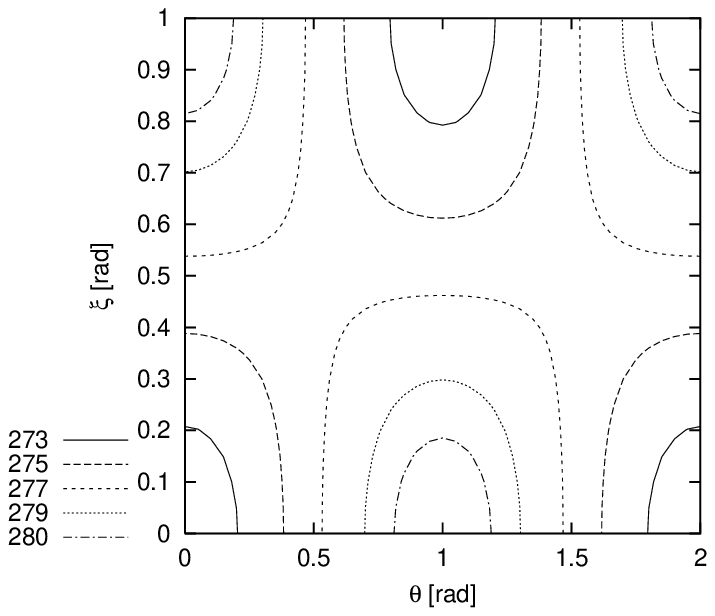}} \hspace{-1.5cm} \epsfxsize 3.5in {\epsfbox{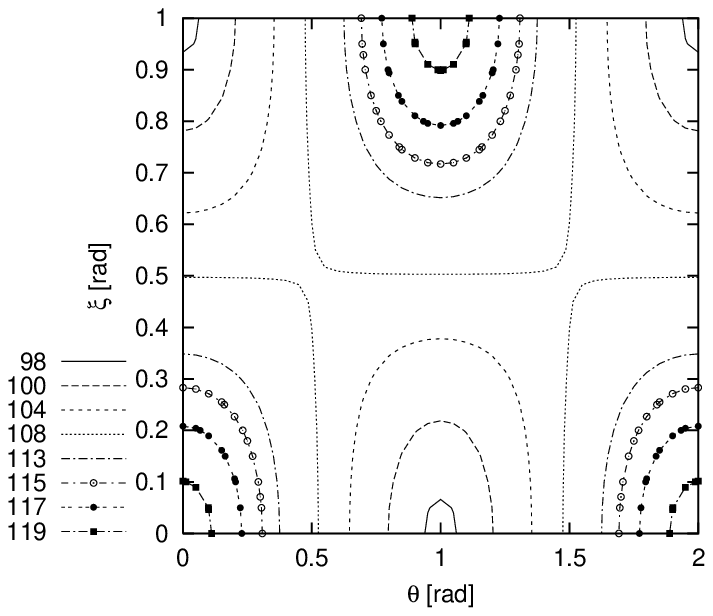}} }
\vskip -0.2in
    \caption{The CP phase dependence of the chargino masses. On the left (right) the heaviest (lightest) chargino mass is depicted in contour plot for the intervals $0\le \theta \le 2\pi$ and $0\le \xi \le \pi$. The values for the contours are given in GeV.}\label{fig:Masses}
\end{figure}

We now  investigate the phase dependence of the masses and take as an example the light chargino mass scenario.  In numerical analyses we assume the following relations  among the parameters of the model: $|M_L|=|M_R|, \, g_R=g_L$, and we set $|M_L|=150$ GeV, $|\mu| = 200$ GeV, and $\tan\beta=10$. As mentioned before, we further assume the VEV's $v_{\Delta_R}$ and $v_{\delta_R}$ to be big enough for ${\tilde \Delta}_R,\; {\tilde \delta}_R$ to decouple from the lighter charginos. In Fig.~\ref{fig:Masses}, the CP-odd phase dependences of the lightest and the heaviest chargino masses are shown in a two-dimensional contour plot. The masses of the four charginos in the CP-conserving case are $\tilde{M}_{\chi^\pm_i}\left(\theta=\xi=0\right)=\left(119.7, 150.0, 200.0, 272.0\right)$ GeV. Among four, the second heaviest is independent of phases and determined completely by the Higgsino mass ($\mu$). This is clear from the chargino mass matrix given in Eq.~(\ref{eq:charmassmatrix}) where the phase dependence of the $\mu$ entry disappears when we consider $M^{c\dagger} M^{c}$. From analyses, we observe that the second lightest mass is practically independent of CP-odd phases so that we only include the figures for the lightest and the heaviest masses.

In the left graph of Fig.~\ref{fig:Masses}, we give the phase dependence of the heaviest chargino mass. This chargino is important for the analysis in Section \ref{sec:decays}.  It can be seen that the dependence is not strong and the mass varies in 272-280 GeV range and the CP-odd phases can enhance its value by at most $3\%$, which occurs in the parameter regions $\left(0^\circ\le\theta\le 36^\circ,144^\circ\le\xi\le 180^\circ\right)$, $\left(144^\circ\le\theta\le 216^\circ, 0^\circ\le\xi\le 36^\circ \right)$, or $\left(244^\circ\le\theta\le 360^\circ,144^\circ\le\xi\le 180^\circ\right)$. These regions are denoted with dashed-dotted line contours in the graph. The ones with the solid line are the regions where the mass takes its lowest value. From the graph, these  are approximately $\left(0^\circ\le\theta,\xi\le 36^\circ\right)$, 
$\left(244^\circ\le\theta\le 360^\circ,0^\circ\le\xi\le 36^\circ\right)$, and $\left(144^\circ\le\theta,\le 216^\circ, 144^\circ\le\xi\le 180^\circ\right)$.

The second graph  in Fig.~\ref{fig:Masses} shows the same phase dependence for the lightest chargino mass $\mathrm{\tilde{M}}_{\chi^{\pm}_{1}}$.  Unlike the case of the heaviest chargino, there is a strong dependence on the CP-odd phases $\theta$ and $\xi$. When we span the whole region in the $\left(\theta,\xi\right)$ space, the mass runs in the interval $(97,120)$ GeV. This means that  up to $20\%$ destructive effects  in the lightest chargino mass could  be generated from CP-odd phases. If one divides the $\xi$ interval into two equal regions and the interval for $\theta$ into 4 equal parts, the mass is at its lowest value for either very small $\xi$ values ($\xi\le 10^\circ$) with $\theta\sim 180^\circ$ or maximum $\xi$ value ($\xi\ge 170^\circ$) with two possible values for $\theta$, either $\theta\le 10^\circ$ or $\theta\ge 170^\circ$. Note that the regions bounded by the contours corresponding to 104 GeV are not experimentally allowed.

\section{\bf Chargino cross sections  in LRSUSY with CP-odd  phases}
\label{sec:production}
At the future Linear Collider one will be able to determine the masses of charginos and the pair production cross section to high accuracies. In this part, we evaluate and illustrate the chargino pair production in unpolarized $e^+e^-$ annihilation with non-trivial phases. The cross sections without CP-violation phases are analyzed in  \cite{Frank:1995dh}. For completeness, we include the expressions for the cross sections here. We introduce the following variables
\begin{eqnarray}
s &=& (q_{1}+q_{2})^{2},\nonumber \\
t &=& (q_{1}-p_{1})^{2},\nonumber \\
u &=& (q_{1}-p_{2})^{2},
\end{eqnarray}
where the momenta of the incoming particles are represented by $q_{1}, q_{2}$ and the momenta of the outgoing particles by $p_{1},p_{2}$. 
The chargino production  occurs at tree-level through $e^{+}e^{-}\rightarrow\gamma,~Z_L,~Z_{R}\rightarrow\chi^{+}_{i}\chi^{-}_{j}$ in the s-channel, and $e^{+}e^{-}\rightarrow\tilde{\nu}_{L,R}\rightarrow\chi^{+}_{i}\chi^{-}_{j}$ in the t-channel. Note that the channels through $Z_R$ and ${\tilde \nu}_R$ are new channels specific to LRSUSY.  The corresponding production cross section and effects of CP-phases have been analyzed in MSSM  \cite{Bartl:2004vi,Choi:1999fs}. 
\begin{figure}[htb]
\vspace{-0.1cm}  
    \centerline{\epsfxsize 5.8in {\epsfbox{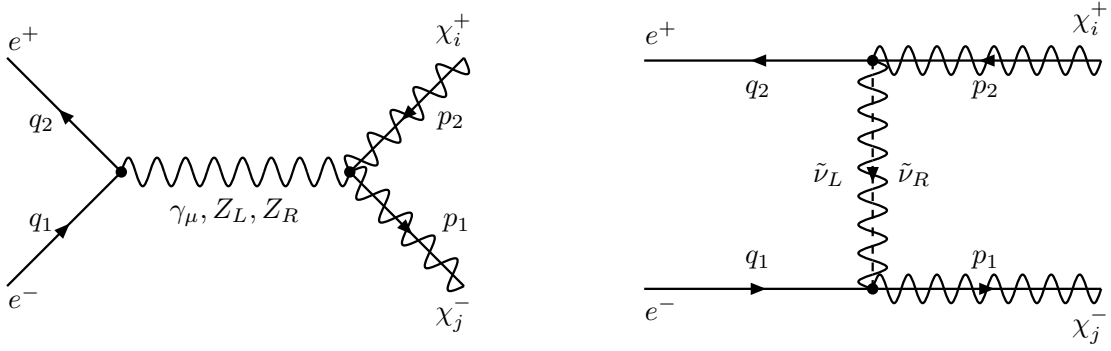}} }
\vskip -0.6cm
    \caption{The s- and t-channel Feynman diagrams contributing to the $e^+e^-\to\chi^+_i\chi^-_j$ scattering.}
\end{figure}

\noindent
The Lagrangian responsible for the interaction is: 
\begin{eqnarray}{\cal L}_{C} &=& -e A_{\mu} \bar{\chi}^{+}_{i}\gamma^{\mu}{\chi}^{+}_{j}+ \frac{g_{L}}{\cos\theta_{W}} Z^{\mu}_{L} \bar{\chi}^{+}_{i}\gamma^{\mu}\biggl[O^{\prime L}_{ij}P_{L}+O^{\prime R}_{ij}P_{R}\biggr]{\chi}^{+}_{j} \nonumber  \\&+&\frac{ g_{R}\sqrt {\cos 2 \theta_{W}}}{\cos\theta_{W}} Z^{\mu}_{L} \bar{\chi}^{+}_{i}\gamma^{\mu}\biggl[O^{\prime L}_{ij}P_{L}+O^{\prime R}_{ij}P_{R}\biggr] {\chi}^{+}_{j} \nonumber \\ 
&-&g \sum_i {\bar \chi}_i^{+ c}
\left \{ \left [V_{i1}^{\ast}P_L-\left ( \frac{m_{\nu_k}U_{i3}}{\sqrt{2}M_W \sin
\beta} +\frac{m_{e_k}U_{i4}}{\sqrt{2}M_W \cos \beta}
\right)P_R\right ]e_k {\tilde \nu}_{Lk}^{\ast} \right.
      \nonumber
\\
      &+&\left. 
\left [ U_{i2}P_R- \left ( \frac{m_{\nu_k}V_{i3}^{\ast}}{\sqrt{2}M_W \sin
\beta} + \frac{m_{e_k}V_{i4}^{\ast}}{\sqrt{2}M_W \cos
\beta}\right )P_L\right ] e_k {\tilde \nu}_{Rk}^{\ast} \right\},
\end{eqnarray}
where $P_{L(R)}=(1\mp\gamma_5)/2$ is the left(right) chirality projection operator and we define the following matrices
\begin{eqnarray}
O^{L}_{ij} &=&V_{i2}V^{\ast}_{j2}+V_{i3}V^{\ast}_{j3}+V_{i4}V^{\ast}_{j4},\nonumber\\O^{R}_{ij} &=&U_{i2}U^{\ast}_{j2}+U_{i3}U^{\ast}_{j3}+U_{i4}U^{\ast}_{j4},\nonumber\\O_{ij}^{\prime L}&=&-V_{i1}V^*_{j1}-\frac{1}{2}V_{i3}V^*_{j3}-\frac{1}{2}V_{i4}V^*_{j4}+\delta_{ij}\sin^2\theta_{\rm W},\nonumber\\
O_{ij}^{\prime R}&=&-U^*_{i1}U_{j1}-\frac{1}{2}U^*_{i3}U_{j3}-\frac{1}{2}U^*_{i4}U_{j4}+\delta_{ij}\sin^2\theta_{\rm W}\nonumber.
\end{eqnarray}
We perform the calculation in the center of mass frame, where the expression for the cross section is given by
\begin{eqnarray}
\frac{d\sigma}{d\cos \theta_{CM}}=\frac{1}{32\,\pi\,s^2}\,\lambda^{1/2}(s,\tilde{M}_{\chi_i}^2,\tilde{M}_{\chi_j}^2)\,|M(s,t)|^{2},
\end{eqnarray}
where we neglect the lepton masses. The triangle function $\lambda$ is defined as $\lambda(x,y,z)\equiv x^2+y^2+z^2-2(x y+x z +y z)$
and $M(s,t)$ is the invariant amplitude of the collision process.  The $Z_H e {\bar e}$ vertex is defined as
\begin{equation}
{\cal L}_{Z_H e {\bar e}}=-\frac{g_{L,R}}{2 \cos \theta_W} {\bar e}\gamma_{\mu} (c_V^{H}-\gamma_5 c_A^{H}) e Z_{H}^{\mu},
\end{equation}
where $H=L,R$, and 
\begin{eqnarray}
&&c_V^L\equiv c_L+c_R=2 \sin ^2 \theta_W-1/2,~~ c_A^L \equiv c_L-c_R=-1/2,\nonumber\\
 &&c_V^R\equiv c^{\prime}_L+c^{\prime}_R= \frac{\sin ^2\theta_W}{\cos 2 \theta_W}-1/2,~~ c_A^R\equiv c_L^{\prime}-c_R^{\prime}=1/2\nonumber.  
\end{eqnarray}

\noindent
The total cross section is obtained by integrating over the angle $\theta_{CM}$. It yields
\begin{eqnarray}
\sigma &=&\sigma_{\gamma}+\sigma_{Z_{L,R}}+\sigma_{{\tilde \nu}_{L}}+\sigma_{{\tilde \nu}_{R}}+\sigma_{\gamma Z_{L,R}}+\sigma_{\gamma {\tilde \nu}_{L, R}} +\sigma_{Z_{L,R } } +\sigma_{Z_{L,R} {\tilde \nu}_{L}}+\sigma_{Z_{L,R} {\tilde \nu}_{R}},
\end{eqnarray}
where the terms in the cross section are as follows
\begin{eqnarray}
\sigma_{\gamma}&&\hspace*{-0.2cm}=\frac{e^{4}}{16\pi s}\delta_{ij}\biggl[1-(y-z)^{2}+\frac{1}{3}\lambda(1,y,z)+4x\biggr]\lambda^{1/2}(1,y,z),
\\
\sigma_{\tilde{\nu}_{L}} && \hspace*{-0.2cm}= \frac{g^{4}}{128\pi s}|V_{i1}|^{2}|V_{j1}|^{2}\biggl\{\biggl[\frac{1}{(a^{2}-b^{2})}-\frac{(y-z)^{2}}{(a^{2}-b^{2})}-4\lambda^{-1/2}(1,y,z)\nonumber\\ &&\times \biggl(\ln(\frac{a+b}{a-b})-\frac{2ab}{a^{2}-b^{2}}\biggr)+4\biggl(1-\frac{a}{b}\ln(\frac{a+b}{a-b})+\frac{a^{2}}{a^{2}-b^{2}}\biggr)\biggr]\biggr\}\lambda^{1/2}(1,y,z),
\\
\sigma_{\tilde{\nu}_{R}} && \hspace*{-0.2cm}= \frac{g^{4}}{128\pi s}|U_{i2}|^{2}|U_{j2}|^{2}\biggl\{\biggl[\frac{1}{(a^{\prime 2}-b^{2})}-\frac{(y-z)^{2}}{(a^{\prime 2}-b^{2})}-4\lambda^{-1/2}(1,y,z)\nonumber\\ 
&&\times \biggl(\ln(\frac{a^{\prime}+b}{a^{\prime}-b})-\frac{2a^{\prime}b}{a^{\prime 2}-b^{2}}\biggr)+4\biggl(1-\frac{a^{\prime}}{b}\ln(\frac{a^{\prime}+b}{a^{\prime}-b})+\frac{a^{\prime 2}}{a^{\prime 2}-b^{2}}\biggr)\biggr]\biggr\}\lambda^{1/2}(1,y,z),
\\
\sigma_{\gamma Z_{L}} && \hspace*{-0.2cm}= \frac{e^{2}g^{2}} {32 \pi\cos^{2}\theta_{W}}\delta_{ij}(c_{L}+c_{R}){\cal R}e[D_{Z_L}(s)(O^{\prime L}_{ij}+O^{\prime R}_{ij})]\nonumber\\ 
&&\times [1-(y-z)^{2}+\frac{1}{3}\lambda(1,y,z)+4x]\lambda^{1/2}(1,y,z),
\\
\sigma_{\gamma Z_{R}} && \hspace*{-0.2cm}= \frac{e^{2}g^{2}} {32\pi\cos^{2}\theta_{W}}\cos2\theta_{W}\delta_{ij}(c_{L}^{\prime}+c_{R}^{\prime}){\cal R}e[D_{Z_R}(s)(O^{L}_{ij}+O^{R}_{ij})] \nonumber\\ 
&&\times [1-(y-z)^{2}+\frac{1}{3}\lambda(1,y,z)+4x]\lambda^{1/2}(1,y,z),
\\
\sigma_{\gamma \tilde{\nu}_{L}} && \hspace*{-0.2cm}= \frac{-e^{2}g^{2}}{64 \pi s}|V_{i1}|^{2}\delta_{ij}\biggl[[1-(y-z)^{2}+4x] \frac{1}{b}\ln(\frac{a+b}{a-b})-4(1+a)\nonumber\\
&&\times \biggl(2-\frac{a}{b}\ln(\frac{a+b}{a-b})\biggr)\biggr]\lambda^{1/2}(1,y,z),
\\
\sigma_{\gamma \tilde{\nu}_{R}} && \hspace*{-0.2cm}= \frac{-e^{2}g^{2}}{64 \pi s}|U_{i2}|^{2}\delta_{ij}\biggl[[1-(y-z)^{2}+4x] \frac{1}{b}\ln(\frac{a^{\prime}+b}{a^{\prime}-b}-4(1+a^{\prime})\nonumber\\
&&\times \biggl(2-\frac{a^{\prime}}{b}\ln(\frac{a^{\prime}+b}{a^{\prime}-b})\biggr)\biggr]\lambda^{1/2}(1,y,z),
\\
\sigma_{Z_{L} \tilde{\nu}_{L}} && \hspace*{-0.2cm}= \frac{-g^{4}}{64 \pi\cos^{2}\theta_{W}}c_L\biggl\{{\cal R}e[D_{Z_L}(s)V_{i1}^{\ast}V_{j1}O^{\prime L}_{ij}] \biggl([1-(y-z)^{2}]\frac{1}{b}\ln(\frac{a+b}{a-b})-4(1+a)\nonumber\\
&&\times [2-\frac{a}{b}\ln(\frac{a+b}{a-b})]\biggr)+{\cal R}e[D_{Z_L}(s)V^{\ast}_{i1}V_{j1}O^{\prime R}_{ij}]\frac{4x}{b}\ln(\frac{a+b}{a-b})\biggr\}\lambda^{1/2}(1,y,z),
\\
\sigma_{Z_{L} \tilde{\nu}_{R}} && \hspace*{-0.2cm}= \frac{-g^{4}}{64 \pi\cos^{2}\theta_{W}}c_R\biggl\{{\cal R}e[D_{Z_L}(s)U_{i2}U^{\ast}_{j2}O^{\prime L}_{ij}] \biggl([1-(y-z)^{2}]\frac{1}{b}\ln(\frac{a^{\prime}+b}{a^{\prime}-b} )- 4(1+a^{\prime})\nonumber\\
&&\times [2-\frac{a^{\prime}}{b}\ln(\frac{a^{\prime}+b}{a^{\prime}-b})]\biggr)+{\cal R}e[D_{Z_L}(s)U_{i2}U^{\ast}_{j2}O^{\prime R}_{ij}]\frac{4x}{b}\ln(\frac{a^{\prime}+b}{a^{\prime}-b})\biggr\}\lambda^{1/2}(1,y,z),
\\
\sigma_{Z_{R} \tilde{\nu}_{L}} && \hspace*{-0.2cm}= \frac{-g^{4}}{64 \pi\cos^{2}\theta_{W}}\cos2\theta_{W} c_L^{\prime}\biggl\{{\cal R}e[D_{Z_R}(s)V_{i1}^{\ast} V_{j1}O^{L}_{ij}] \biggl([1-(y-z)^{2}] \frac{1}{b}\ln(\frac{a+b}{a-b})\nonumber\\
&&-4(1+a)[2-\frac{a}{b}\ln(\frac{a+b}{a-b})]\biggr)+{\cal R}e[D_{Z_R}(s)V^{\ast}_{i1}V_{j1}O^{R}_{ij}]\frac{4x}{b} \ln(\frac{a+b}{a-b})\biggr\}\nonumber \\
&&\times \lambda^{1/2}(1,y,z),
\\
\sigma_{Z_{R} \tilde{\nu}_{R}} && \hspace*{-0.2cm}= \frac{-g^{4}}{64 \pi\cos^{2}\theta_{W}}\cos2\theta_{W} c_R^{\prime}\biggl\{{\cal R}e[D_{Z_R}(s)U_{i2}U^{\ast}_{j2}O^{L}_{ij}] \biggl([1-(y-z)^{2}] \frac{1}{b}\ln(\frac{a^{\prime}+b}{a^{\prime}-b})\nonumber\\
&&-4(1+a^{\prime})[2-\frac{a^{\prime}}{b}\ln(\frac{a^{\prime}+b}{a^{\prime}-b})]\biggr)+{\cal R}e[D_{Z_R}(s)U_{i2}U^{\ast}_{j2}O^{R}_{ij}]\frac{4x}{b} \ln(\frac{a^{\prime}+b}{a^{\prime}-b})\biggr\}\nonumber \\&&\times \lambda^{1/2}(1,y,z),
\\
\sigma_{Z_{L,R}} && \hspace*{-0.2cm}= \frac{g^{4}s}{64 \pi\cos^{4}\theta_{W}}\cos2\theta_{W} (c_{L}c_L^{\prime}+ c_{R}c_R^{\prime} ) \biggl[
{\cal R}e[D_{Z_L}(s)D_{Z^{\ast}_R} (s)( O^{\prime L}_{ij}O^{L \ast }_{ij}+O^{\prime R}_{ij}O^{R \ast }_{ij})] \nonumber \\
&&\times [1-(y-z)^{2}+\frac{1}{3}\lambda(1,y,z)]+ 4x{\cal R}e[D_{Z_L}(s)D_{Z^{\ast}_R}(s)(O^{\prime L}_{ij}O^{R \ast}_{ij}+O^{L \ast}_{ij}O^{\prime R}_{ij})] \biggr] \nonumber \\
&& \times ]\lambda^{1/2}(1,y,z),
\\
\sigma_{Z_L} && \hspace*{-0.2cm}= \frac{g^{4}s}{64 \pi\cos^{4}\theta_{W}}|D_{Z_L}(s)|^{2}(c_{L}^{2}+c_{R}^{2}) \biggl\{(|O^{\prime L}_{ij}|^{2}+|O^{\prime R}_{ij}|^{2}) \nonumber \\  
&&\times [1-(y-z)^{2}+\frac{1}{3}\lambda(1,y,z)]+4x{\cal R}e[{O}^{\prime L}_{ij}\,{O}^{\prime R \ast}_{ij}] \biggr\} \lambda^{1/2}(1,y,z),
\\
\sigma_{Z_R} && \hspace*{-0.2cm}= \frac{g^{4}s \cos^{2}2\theta_{W}}{64 \pi\cos^{2}\theta_{W}}|D_{Z_R}(s)|^{2}(c_{L}^{\prime 2}+c_{R}^{\prime 2})\biggl\{(|O^{L}_{ij}|^{2}+|O^{R}_{ij}|^{2})
\nonumber \\ 
&&\times [1-(y-z)^{2}+\frac{1}{3}\lambda(1,y,z)]+4x{\cal R}e [O^{L}_{ij}\,O^{R \ast}_{ij}] \biggr\}\lambda^{1/2}(1,y,z).
\end{eqnarray}
Here, we have taken $g_L(=g_R) \equiv g$, $a =-(1-y-z)/2-{M}^{2}_{{\tilde \nu}_{L}}/s ; \;a^{\prime}= -(1-y-z)/2-{M}^{2}_{{\tilde \nu}_{R}}/s$ and $b=(1/2)\lambda^{1/2}(1,y,z)$; 
\noindent
 $D_{Z_{L,R}}(s)=(s-M^{2}_{Z}+iM_{Z}\Gamma_{Z})^{-1}_{L,R}$, 
 and the variables $x,y,z$ are defined as
\begin{eqnarray}
x = \frac{\tilde{M}_{\chi^{\pm}_{i}}\tilde{M}_{\chi^{\pm}_{j}}}{s},\;\;\;\;\;
y = \frac{(\tilde{M}_{\chi^{\pm}_{i}})^{2}}{s},\;\;\;\;\;
z = \frac{(\tilde{M}_{\chi^{\pm}_{j}})^{2}}{s}.
\end{eqnarray}

\begin{figure}[h]
\vspace{-0.05in}  
    \centerline{\hspace*{3.1cm} \epsfxsize 3.5in {\epsfbox{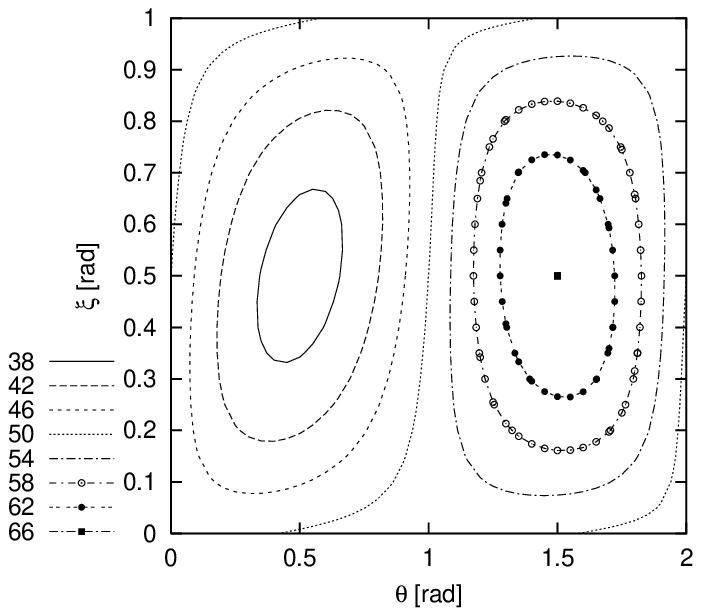}} \hspace{-1.5cm} \epsfxsize 3.5in {\epsfbox{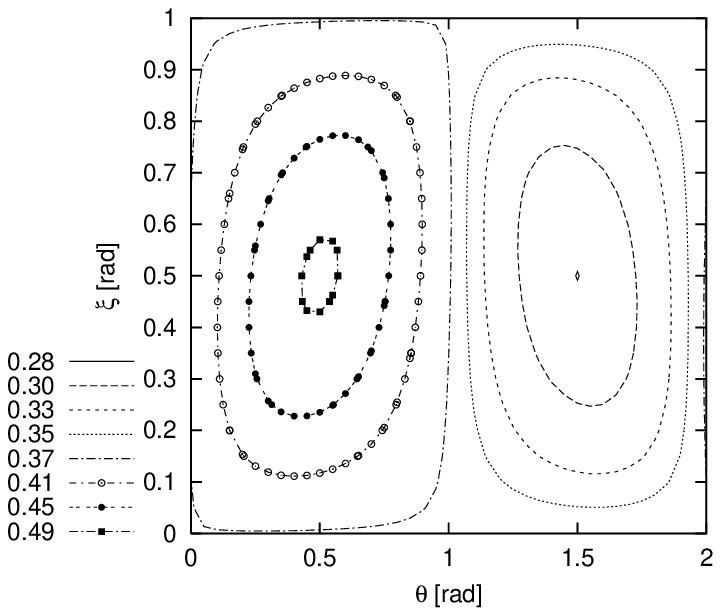}} }
\vskip -0.2in
    \caption{The CP phase dependence of the cross section of the process $e^+e^-\to\chi^+_1\chi^-_1$. The left (right) figure is at $\sqrt{s}=500 (1000)$ GeV. All the other parameters of the model are chosen as stated in Section~\ref{sec:masses}. Also, $M_{Z_R}=500$ GeV, $\Gamma_{Z_R}=20$ GeV, $M_{\tilde{\nu}_L}=150$ GeV and $M_{\tilde{\nu}_R}=600$ GeV. The values for the contours are given in pb.}\label{fig:cs11}
\end{figure}

\begin{figure}[htb]
\vspace{-0.05in}  
    \centerline{\hspace*{3.1cm} \epsfxsize 3.5in {\epsfbox{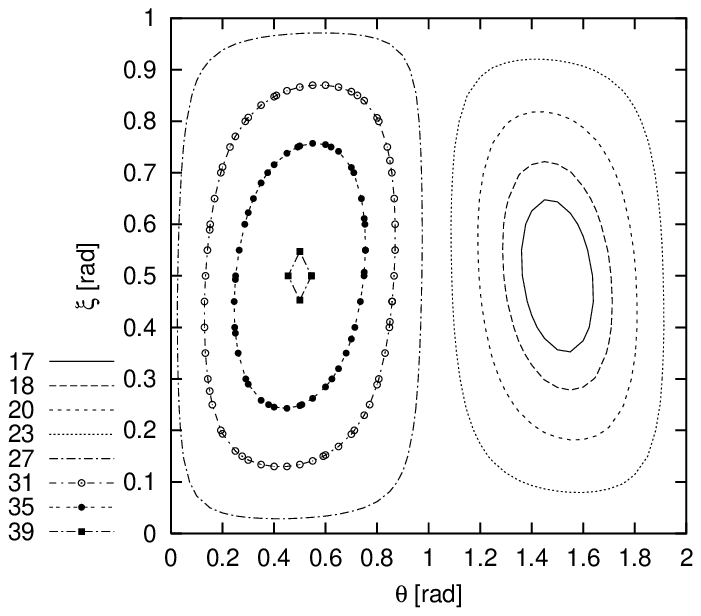}} \hspace{-1.5cm} \epsfxsize 3.5in {\epsfbox{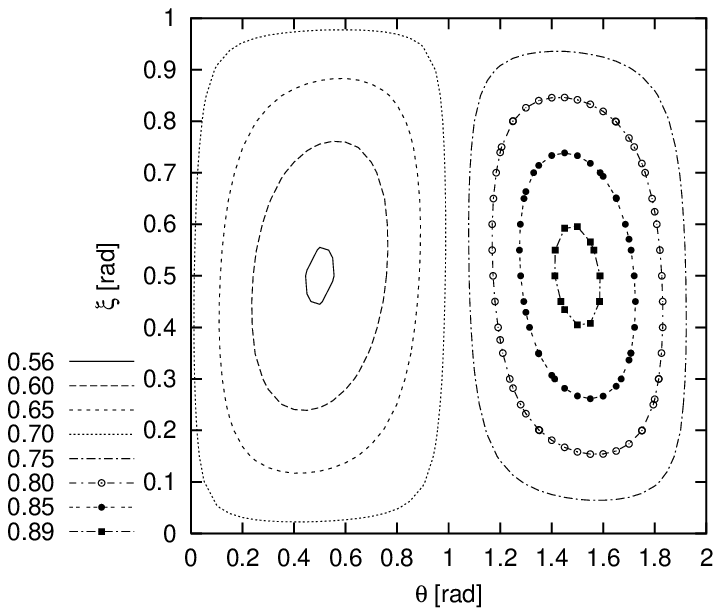}} }
\vskip -0.2in
    \caption{Same as Fig.~\ref{fig:cs11} but for $e^+e^-\to\chi^+_2\chi^-_2$.}\label{fig:cs22}
\end{figure}

In Fig.~\ref{fig:cs11}, the cross section for the process $e^+e^-\to \chi^+_1\chi^-_1$ in the ($\theta, \xi$)-plane is depicted at two different center of mass energies, 500 and 1000 GeV. The value of the cross section without CP-odd phases is around 50 pb for $\sqrt{s} =500$ GeV and 0.35 pb for $\sqrt{s} =1000$ GeV. CP-odd phases can lead to either constructive or destructive interference effects, about a maximum $32\%$ for constructive and $25\%$ for destructive effects at $\sqrt{s} =500$ GeV. These occur at $(3\pi/2, \pi/2)$ and $(\pi/2, \pi/2)$ in the $(\theta,\xi)$-plane, respectively. For $\sqrt{s} =1000$ GeV, there exists a similar situation with about $40\%(20\%)$ constructive (destructive) effects. In this case, the maximum occurs at $(\pi/2, \pi/2)$. The case with final states  $\chi^+_2\chi^-_2$
is shown in Fig.~\ref{fig:cs22}. In this case, constructive (destructive) effect can reach up to $60\%(30\%)$ at $\sqrt{s} =500$ GeV and $30\%(20\%)$ at $\sqrt{s} =1000$ GeV.  
\begin{figure}[htb]
\vspace{-0.05in}  
    \centerline{\hspace*{3.1cm} \epsfxsize 3.5in {\epsfbox{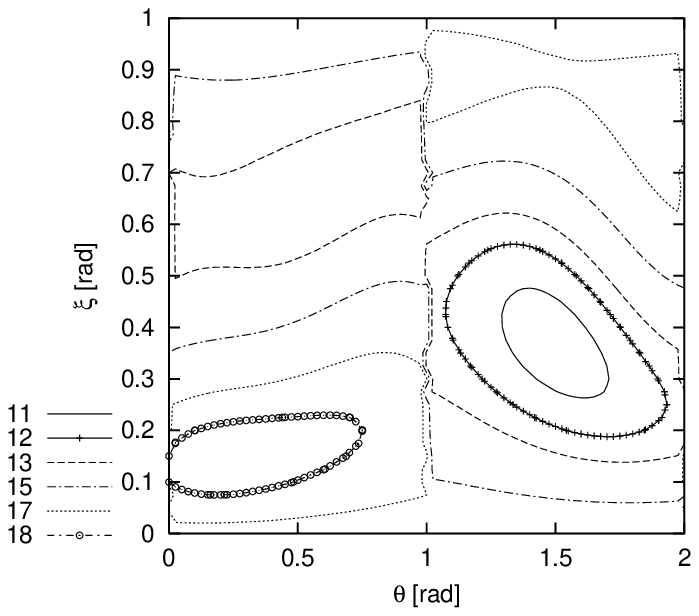}} \hspace{-1.5cm} \epsfxsize 3.5in {\epsfbox{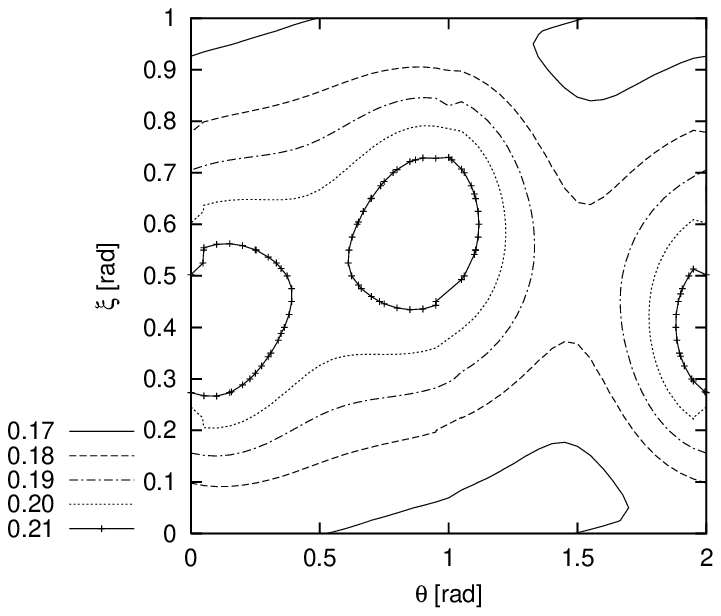}} }
\vskip -0.2in
    \caption{Same as Fig.~\ref{fig:cs11} but for $e^+e^-\to\chi^+_1\chi^-_2$.}\label{fig:cs12}
\end{figure}

\begin{figure}[t]
\vspace{-0.05in}  
    \centerline{\hspace*{3.1cm} \epsfxsize 3.5in {\epsfbox{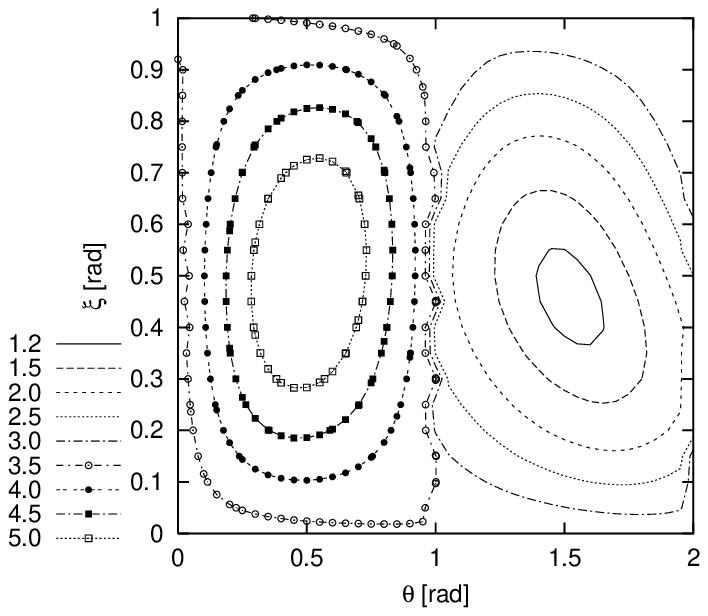}} \hspace{-1.5cm} \epsfxsize 3.5in {\epsfbox{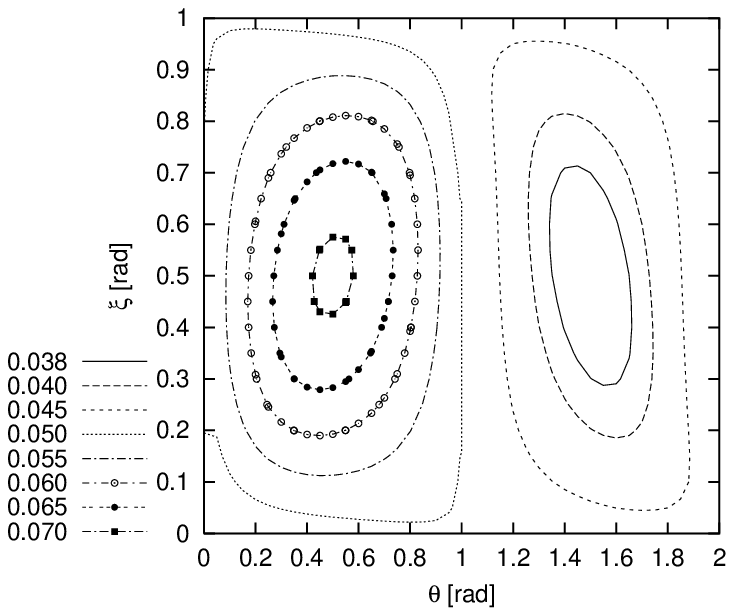}} }
\vskip -0.2in
    \caption{Same as Fig.~\ref{fig:cs11} but for $e^+e^-\to\chi^+_1\chi^-_4$.}\label{fig:cs14}
\end{figure}

Figs.~\ref{fig:cs12} and \ref{fig:cs14} represent the CP-odd phase dependence of the cross sections for $e^+e^-\to\chi^+_1\chi^-_2$ and $e^+e^-\to\chi^+_1\chi^-_4$, respectively. The extrema for the cross section occur in more than one region in the $(\theta,\xi)$ plane. For the $\chi^+_1\chi^-_2$ case at $\sqrt{s} =500$ GeV the maximum cross section is obtained for $\theta\le 3\pi/4$ and $\xi\in[\pi/10,\pi/5]$, and the minimum at around $3\pi/4\le \theta \le 7\pi/4$ and $\pi/4\le\xi\le \pi/2$. The CP-odd phase effects are slightly smaller for  constructive interference:  around $10\%(30\%)$ for the constructive (destructive) case. At $\sqrt{s} =1000$ GeV, there is almost no destructive interference in the entire plane. The cross section can get enhanced at most by $20\%$ in three different intervals shown in Fig.~\ref{fig:cs12}. The $\chi^+_1\chi^-_4$ case is similar to the $\chi^+_1\chi^-_1$ or $\chi^+_2\chi^-_2$ and the  $\theta$ values for getting a maximum (minimum) value for the cross section are the same (around $\pi/2$ for constructive and $3\pi/2$ for destructive) each at different center of mass energies, 500 and 1000 GeV. The enhancement can be more than $50\%$ for both energies.  The destructive effects cannot exceed $20\%$ at $\sqrt{s} =1000$ GeV, but become more than $60\%$ at $\sqrt{s} =500$ GeV. 
\begin{figure}[t]
\vspace{-0.05in}  
    \centerline{\hspace*{3.1cm} \epsfxsize 3.5in {\epsfbox{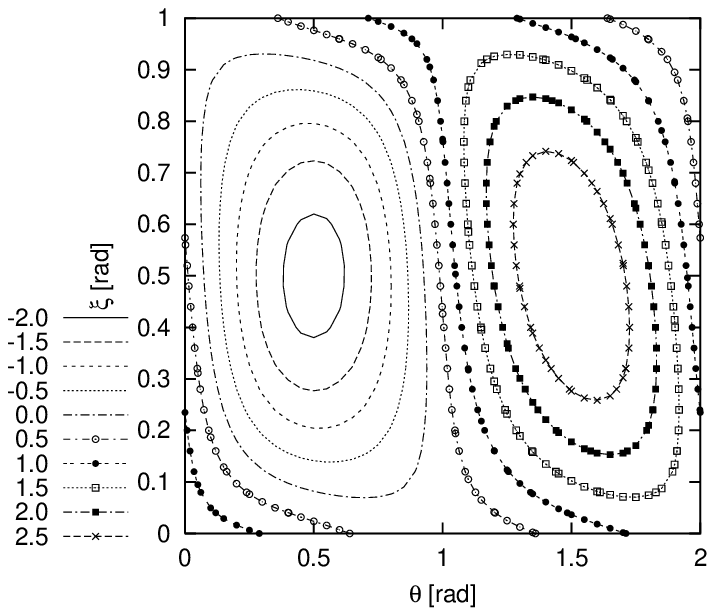}} \hspace{-1.5cm} \epsfxsize 3.5in {\epsfbox{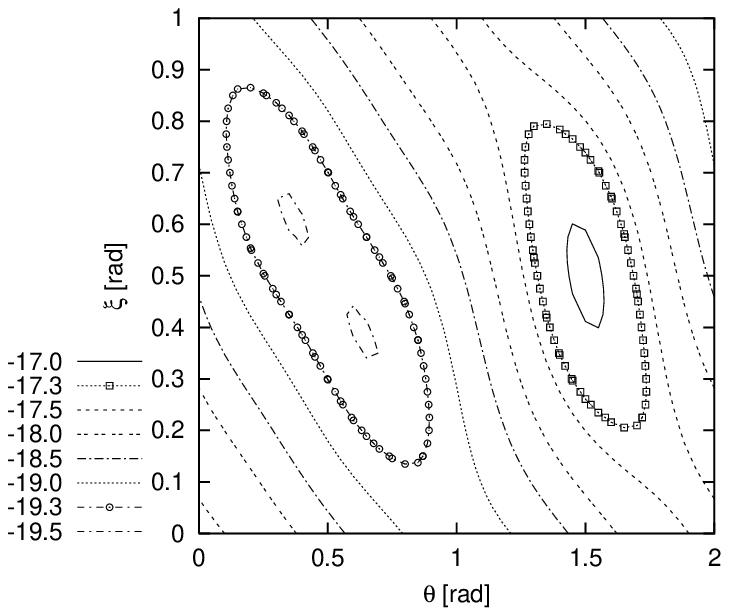}} }
\vskip -0.2in
    \caption{The CP phase dependence of the forward-backward asymmetry ($A_{FB}$) in the process $e^+e^-\to\chi^+_1\chi^-_1$. The left (right) figure is at $\sqrt{s}=500 (1000)$ GeV. All the other parameters of the model are chosen as stated in Section~\ref{sec:masses}. Also, $M_{Z_R}=500$ GeV, $\Gamma_{Z_R}=20$ GeV, $M_{\tilde{\nu}_L}=150$ GeV and $M_{\tilde{\nu}_R}=600$ GeV. The values for the contours are given in percentage.}\label{fig:AFB11}
\end{figure}
We also consider the angular distribution $d\sigma/d\cos \theta_{CM}$ where $ \theta_{CM}$ is angle between the chargino and the electron beam, as well as the forward backward asymmetry ($A_{FB}$) defined as:
\begin{equation}
A_{FB}=\frac{\displaystyle \int \limits^{1}_{0} \left (\frac{d \sigma}{d \cos  \theta_{CM}}\right ) d\cos \theta_{CM}-\int \limits^0_{-1}\left (\frac{d \sigma}{d \cos  \theta_{CM}} \right )d\cos \theta_{CM}}{\displaystyle \int \limits^1_{0}\left (\frac{d \sigma}{d \cos  \theta_{CM}} \right )d\cos  \theta_{CM}+\int \limits^0_{-1}\left (\frac{d \sigma}{d \cos  \theta_{CM}}\right ) d\cos  \theta_{CM}}\,.
\end{equation}
\begin{figure}[htb]
\vspace{-0.05in}  
    \centerline{\hspace*{3.1cm} \epsfxsize 3.5in {\epsfbox{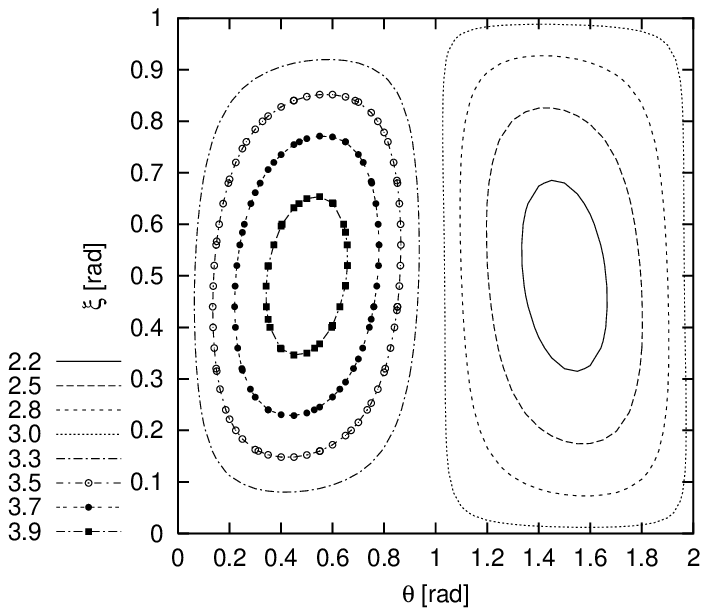}} \hspace{-1.5cm} \epsfxsize 3.5in {\epsfbox{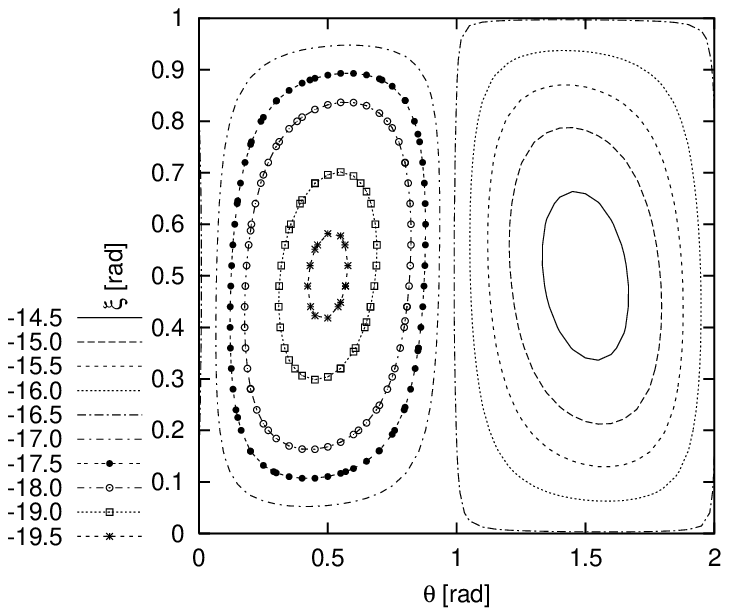}} }
\vskip -0.2in
    \caption{Same as Fig.~\ref{fig:AFB11} but for $e^+e^-\to\chi^+_2\chi^-_2$.}
    \label{fig:AFB22}
\end{figure}
\begin{figure}[htb]
\vspace{-0.05in}  
    \centerline{\hspace*{3.1cm} \epsfxsize 3.5in {\epsfbox{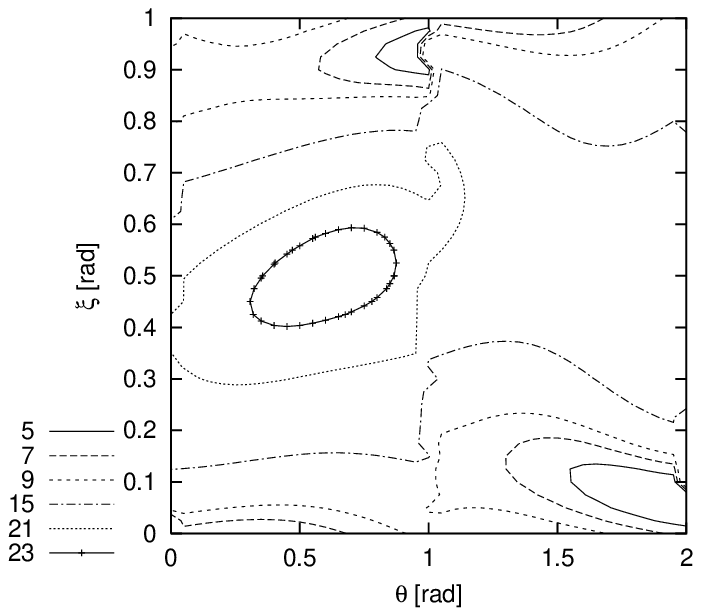}} \hspace{-1.5cm} \epsfxsize 3.5in {\epsfbox{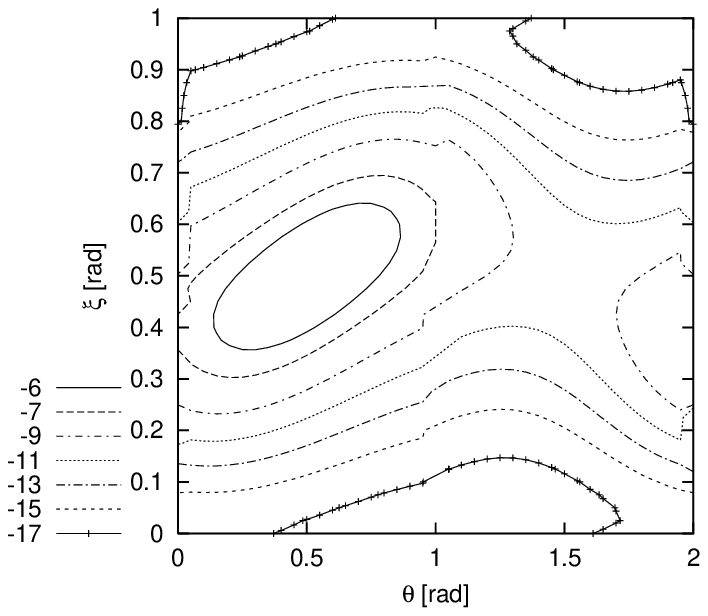}} }
\vskip -0.2in
    \caption{Same as Fig.~\ref{fig:AFB11} but for $e^+e^-\to\chi^+_1\chi^-_2$.}
    \label{fig:AFB12}
\end{figure}
In Fig.~\ref{fig:AFB11}, the forward-backward asymmetry ($A_{FB}$) is shown in the $(\theta,\xi)$-plane for $e^+e^-\to\chi^+_1\chi^-_1$ at both $\sqrt{s} =500$ GeV (on the left) and $\sqrt{s} =1000$ GeV (on the right). The asymmetry without CP-odd phases is around $1.2\%$ at $\sqrt{s} =500$ GeV and $17\%$ at $\sqrt{s} =1000$ GeV. Even though the asymmetry is a lot bigger at $\sqrt{s} =1000$ GeV than at $\sqrt{s} =500$ GeV, it is more sensitive to the CP-odd phases for latter case. The change in $A_{FB}$ can be as much as $50\%$ and can even change sign (there is an almost symmetric pattern with respect to $\theta=\pi$ line). However, the increase in $A_{FB}$ can not exceed $10\%$ at $\sqrt{s} =1000$ GeV. For $\theta=3\pi/2$ and $\xi=\pi/2$, CP-odd phases effects cancel each other. Maximum constructive effects occur around $\pi/2$ for both angles. As depicted in Fig.~\ref{fig:AFB22}, the maximum asymmetry for $e^+e^-\to\chi^+_2\chi^-_2$ is obtained at around $\theta,\xi\sim\pi/2$ at both  $\sqrt{s} =500$ and $1000$ GeV (around $30\%$ and $20\%$ increase, respectively). Maximum destructive interference effects are slightly smaller and take place for $\theta\sim3\pi/2, \xi\sim\pi/2$. 
The final set of graphs for $A_{FB}$ is shown in Fig.~\ref{fig:AFB12} for the process $e^+e^-\to\chi^+_1\chi^-_2$. At $\sqrt{s} =500$ GeV, CP-odd phases have constructive effects in almost the entire range and the asymmetry can reach a maximum $23\%$, for  which $A_{FB}$ is about $360\%$   
bigger than the case without CP phases. For values of the phases around $\pi/2$, destructive effects get maximized at $\sqrt{s} =1000$ GeV and suppress the $A_{FB}$ to be as small as $6\%$ for the phases around $\pi/2$. The suppression in percentage is almost equal to the enhancement at $\sqrt{s} =500$ GeV.

\section{Chargino Decays in LRSUSY with CP violating phases}
\label{sec:decays}
In this section we consider chargino production by a polarized $e^+e^-$ beam, followed by two body chargino decay. We calculate the associated production and decay cross section and the observable T-odd asymmetry.

Two-body decays of charginos have been studied in the context of the MSSM without CP violating phases \cite{Gunion:1987yh}. There, the rates of interest are for decays to SM particles: $ \chi^+_i \rightarrow \chi_j^0 W_L^+,\; \chi_i^+ \rightarrow \chi_j^+ Z_L$, if kinematically allowed. Though these processes are also allowed in LRSUSY, their decay ratios do not differ significantly from the ones in MSSM. We choose to concentrate here on those decays which could show a difference in signals between MSSM and LRSUSY. These are the decays to leptons/sleptons, which are sensitive to the right-handed gauginos. These decay widths are:
\begin{eqnarray}
\label{decays}
\Gamma_D (\chi^+_i \rightarrow l_j^+{\tilde \nu}_{Lj})&=& \frac{g^2{\tilde M}_{\chi_i^+} \lambda^{1/2}(1, y^2_l, y^2_{{\tilde \nu}_L})}{64 \pi y_W^2 \cos^2 \beta} \left [ \left (1+y_l^2-y_{{\tilde \nu}_L}^2 \right )\left (y_l^2 |U_{i4}|^2+2y_W^2 \cos^2\beta|V_{i1}|^2 \right ) \right.
\nonumber \\
&-&\left. 4\sqrt{2} y_W y_l \cos \beta\; {\cal R}e[V_{i1}U_{i4}] \right], \nonumber \\
\Gamma_D (\chi^+_i \rightarrow l_j^+{\tilde \nu}_{Rj})&=& \frac{g^2{\tilde M}_{\chi_i^+} \lambda^{1/2}(1, y^2_l, y^2_{{\tilde \nu}_R})}{64 \pi y_W^2 \cos^2 \beta} \left [ \left (1+y_l^2-y_{{\tilde \nu}_R}^2 \right ) \left (y_l^2 |V_{i4}|^2+2y_W^2 \cos^2\beta|U_{i2}|^2 \right ) \right.
\nonumber \\
&-&\left. 4\sqrt{2} y_W y_l \cos \beta \;{\cal R}e[U_{i2}V_{i4}] \right],  \\
\Gamma_D (\chi^+_i \rightarrow {\tilde l}_{Lj}^+{ \nu}_{j})&=& \frac{g^2{\tilde M}_{\chi_i^+} \lambda^{1/2}(1, y^2_{{\tilde l}_L}, 0)}{64 \pi y_W^2 \cos^2 \beta} \left [ \left (1+y_l^2-y_{{\tilde l}_L}^2 \right ) \left (y_l^2 |V_{i4}|^2+2y_W^2 \cos^2\beta|U_{i1}|^2 \right ) \right ],
\nonumber \\
\Gamma_D (\chi^+_i \rightarrow {\tilde l}_{Rj}^+{ \nu}_{j})&=& \frac{g^2{\tilde M}_{\chi_i^+} \lambda^{1/2}(1, y^2_{{\tilde l}_R}, 0)}{64 \pi y_W^2 \cos^2 \beta} \left [ \left (1+y_l^2-y_{{\tilde l}_R}^2 \right ) \left (y_l^2 |U_{i4}|^2+2y_W^2 \cos^2\beta|V_{i2}|^2 \right ) \right ], \nonumber
\end{eqnarray}
where  $y_{l; {\tilde \nu}_L; {\tilde \nu}_R; {\tilde l}_L; {\tilde l}_R; W}\!\!=\!\! \frac {M_{l; {\tilde \nu}_L; {\tilde \nu}_R; {\tilde l}_L; {\tilde l}_R; W}}{{\tilde M}_{\chi_i}}$. 
Similarly, charginos could decay to quarks/squarks, if kinematically allowed, though if sleptons are lighter, we expect the decay rates to (s)leptons to dominate over the ones to (s)quarks.
In principle the decay processes are similar, and we have, as in the MSSM \cite{Gunion:1987yh}:
\begin{equation}
{\rm BR} (\chi^+ \rightarrow {\rm all~ squarks})\sim 3\;{\rm BR} (\chi^+ \rightarrow {\rm all~ sleptons})
\end{equation}
if kinematically allowed. These types of decays would be less important phenomenologically than the decays into neutralinos if the lightest neutralino is the lightest supersymmetric particle (LSP). The first two would be dominant if the sneutrino is the LSP and the last two would be possible for a heavier chargino if ${\tilde M}_{\chi_i}> M_{\tilde l_j} $, and if the neutralino is the LSP. In the case the neutralino is the LPS, the scalar leptons and scalar neutrinos would decay further, $\chi_i^+ \rightarrow \chi^0_j l^+ \nu$. We concentrate in this section on the first two decay channels, since we expect these to show the strongest deviations from MSSM (due to the availability of the right-handed sneutrino).

We study the chargino production $e^+ e^- \to \chi^+_i \chi^-_j$ with longitudinally polarized beams, and the subsequent decay of one of the charginos into a sneutrino and anti-lepton $\chi_i^+ \to l^+ {\tilde \nu}_{L,R}$. We define the triple product:
\begin{eqnarray}
{\cal T}_l=(\vec {p}_{e^-}\times \vec{p}_{\chi_i^+})\cdot \vec{p}_{l^+}
\end{eqnarray}
and the T-odd asymmetry:
\begin{eqnarray}
{\cal A}_l^T= \frac{\displaystyle\sigma_{PD}({\cal T}_l >0)-\sigma_{PD}({\cal T}_l<0)}{\sigma_{PD}(\displaystyle{\cal T}_l >0)+\sigma_{PD}({\cal T}_l<0)}\,,
\end{eqnarray}
where here the cross section $\sigma_{PD}$ is the product of the chargino pair production cross section and the branching ratio for the decay into the sneutrino. This asymmetry is sensitive both to the phases in the chargino mass matrix, and to absorptive contributions, which do not contribute to CP violation. To eliminate the absorptive effects, we study the CP asymmetry:
\begin{eqnarray}
{\cal A}_l= \frac12 \left ({\cal A}_l^T- \bar{\cal A}_l^T\right )
\end{eqnarray}
with $ \bar{\cal A}_l^T$ the asymmetry for the CP conjugated process  $e^+ e^- \to \chi^-_i \chi^+_j,\; \chi_i^- \to l^- \bar {\tilde \nu}_{L,R}$.

In order to calculate the cross section for the combined process of chargino production and decay of $\chi^+_i$, we use the spin density matrix formalism for the production and decays of charginos as in  \cite{Kittel:2004kd}.  The amplitude for the whole process is:
\begin{eqnarray}
T= 2\Delta(\chi^+_i) \sum \limits_{\lambda_i} T_P^{\lambda_i}T_{D, \lambda_i},
\end{eqnarray}
where $T_P^{\lambda_i}$ is the helicity amplitude for the production, $T_{D, \lambda_i}$ the helicity amplitude for the decay, and $\Delta(\chi^{+}_{i})=1/(s_i-{\tilde M}^2_{\chi_{i}}+i{\tilde M}_{\chi_{i}}\Gamma_{\chi_{i}})$ is the chargino propagator. In the above formulas $\lambda_{i}$ is the helicity of the decaying chargino, and $s_i$ its four-momentum squared.
The total differential cross section is defined as:
\begin{eqnarray}
d \sigma_{PD}=\frac{1}{8E_b^2}|T|^2 (2 \pi)^4 \delta^4\left ( p_{e^+}+p_{e^-}-\sum \limits_i p_{i} \right)\prod \limits_i\frac{d^3 p_{i}}{(2 \pi)^3 2p_i^0}\,,
\end{eqnarray}
where the index $i$ runs over the charginos, sneutrino and emitted lepton momenta and $E_b$ is the energy of the incoming beam.

In the definition of the asymmetry:
\begin{eqnarray}
{\cal A}_l^T= \frac {\displaystyle \int {\rm Sign} {\cal T}_l |T|^2 \delta^4\left ( p_{e^+}+p_{e^-}-\sum \limits_i p_{i} \right)\prod \limits_i \frac {d^3 p_{i}} {(2 \pi)^3 2p_i^0} } {\displaystyle\int  |T|^2\delta^4\left ( p_{e^+}+p_{e^-}-\sum \limits_i p_{i} \right) \prod \limits_i \frac {d^3 p_{i}} {(2 \pi)^3 2p_i^0} }\,, 
\end{eqnarray}
the only CP sensitive contribution comes from the production and decay amplitudes corresponding to $\chi_i^+$ polarization perpendicular to the production plane. In the notation of \cite{Kittel:2004kd}, the production amplitudes are:
\begin{eqnarray}
\Sigma^2_P&=& \Sigma^2_P(Z_LZ_L)+ \Sigma^2_P(Z_LZ_R)+\Sigma^2_P(Z_RZ_R)+\Sigma^2_P({\tilde \nu}_LZ_L)
+\Sigma^2_P({\tilde \nu}_RZ_L)\nonumber\\
&+&\Sigma^2_P({\tilde \nu}_LZ_R)+\Sigma^2_P({\tilde \nu}_RZ_R)
\end{eqnarray}
with
\begin{eqnarray}
 \Sigma^2_P(Z_LZ_L)&=& 2 \frac{g^4}{\cos^4 \theta_W} |D_{Z_L}|^2 \left [c_R^2(1+P_{e^-})(1-P_{e^+})-c_L^2(1-P_{e^-})(1+P_{e^+}) \right ] \nonumber \\
 &&\times E_b^2M_{\chi_i^-} q \sin \Theta \; {\cal I}m \left (O^{\prime L}_{ij} O^{\prime \ast R}_{ij}\right),
 \nonumber \\
 \Sigma^2_P(Z_LZ_R)&=& \frac{g^4\cos 2 \theta_W}{\cos^4 \theta_W}  \left [c_Rc_R^{\prime}(1+P_{e^-})(1-P_{e^+})-c_Lc_L^{\prime}(1-P_{e^-})(1+P_{e^+}) \right ] \nonumber \\
 &&\times E_b^2M_{\chi_i^-} q \sin \Theta \; {\cal I}m \left ( D_{Z_L}D_{Z_R}^{\ast}O^{\prime L}_{ij} O^{ \ast R}_{ij}+ D^{\ast}_{Z_L}D_{Z_R}O^{ L}_{ij} O^{ \prime \ast R}_{ij}\right),
  \nonumber \\
 \Sigma^2_P(Z_RZ_R)&=& 2 \frac{g^4\cos^2 2 \theta_W}{\cos^4 \theta_W} |D_{Z_R}|^2 \left [c_R^{\prime 2}(1+P_{e^-})(1-P_{e^+})-c_L^{\prime 2}(1-P_{e^-})(1+P_{e^+}) \right ] \nonumber \\
 &&\times E_b^2M_{\chi_i^-} q \sin \Theta \; {\cal I}m \left (O^{ L}_{ij} O^{ \ast R}_{ij}\right),
 \nonumber \\
 \Sigma^2_P({\tilde \nu}_LZ_L)&=& \frac{g^4}{\cos^2 \theta_W} c_L^2(1-P_{e^-})(1+P_{e^+}) E_b^2M_{\chi_i^-} q \sin \Theta
 \; {\cal I}m \left (V_{i1}^{\ast}V_{j1} O^{\prime R}_{ij} D_{Z_L} D_{{\tilde \nu}_L}^{\ast} \right),
 \nonumber \\ 
\Sigma^2_P({\tilde \nu}_R Z_L)&=&  \frac{g^4}{\cos^2 \theta_W} c_R^2(1+P_{e^-})(1-P_{e^+}) E_b^2M_{\chi_i^-} q \sin \Theta
 \; {\cal I}m \left (U_{i2}U_{j2}^{\ast} O^{\prime L}_{ij} D_{Z_L} D_{{\tilde \nu}_R}^{\ast} \right),
\nonumber \\ 
\Sigma^2_P({\tilde \nu}_L Z_R)&=& \frac{g^4 \cos 2 \theta_W}{\cos^2 \theta_W} c_L^{\prime 2}(1-P_{e^-})(1+P_{e^+}) E_b^2M_{\chi_i^-} q \sin \Theta
 \; {\cal I}m \left (V_{i1}^{\ast}V_{j1} O^{ R}_{ij} D_{Z_R} D_{{\tilde \nu}_L}^{\ast} \right),
 \nonumber \\ 
\Sigma^2_P({\tilde \nu}_RZ_R)&=& \frac{g^4 \cos 2 \theta_W}{\cos^2 \theta_W} c_R^{\prime 2}(1+P_{e^-})(1-P_{e^+}) E_b^2M_{\chi_i^-} q \sin \Theta
 \; {\cal I}m \left (U_{i2}U_{j2}^{\ast} O^{ L}_{ij} D_{Z_R} D_{{\tilde \nu}_R}^{\ast} \right).
 \nonumber \\
\end{eqnarray}
In these expressions the angle $\Theta$ is the scattering angle between the incoming electron beam  and the $\chi_j^-$ chargino in the laboratory frame,  $E_b=\sqrt{s}/2$ is the beam energy, $P_{e^{+(-)}}$ is the positron (electron) polarization,  and $q=E_b\lambda^{1/2}(1, y, z)$. Similarly, the chargino decay amplitudes are:
 \begin{eqnarray}
&&\Sigma^2_D(l^+ {\tilde \nu}_L)=-g^2 \left (|V_{i1}|^2-\frac {m_l^2}{2M_W^2 \cos^2\beta}|U_{i4}|^2 \right ) {\tilde M}_{\chi^+_i} ( {s}_{\chi^+_i}^{(2)} \cdot {p}_{l^+}),
 \nonumber \\
&&\Sigma^2_D(l^+ {\tilde \nu}_R)=-g^2 \left (|U_{i2}|^2-\frac {m_l^2}{2M_W^2 \cos^2\beta}|V_{i4}|^2\right) {\tilde M}_{\chi^+_i} ( {s}_{\chi^+_i}^{(2)} \cdot {p}_{l^+}), 
\end{eqnarray}
and the conjugated decay process $\chi_i^- \to l^- \bar{\tilde \nu}$ amplitudes have an overall $+$ sign.  Here $s_{\chi_i^+}^{(2)}$ is the spin polarization component perpendicular to the production plane and $p_{l^+}$ is the momentum of the antilepton. See  \cite{Kittel:2004kd} for details. The production contribution independent of the chargino polarization vectors can be obtained from the expressions for the cross sections in Section \ref{sec:production}, in the same way as the polarization-independent part of the chargino decay can be obtained from the partial decay widths, Eq. (\ref{decays}). For example:
 \begin{eqnarray}
D(l^+ {\tilde \nu}_L)\hspace*{-0.2cm}&&=\frac{g^2}{2}\left [ \left (|V_{i1}|^2-\frac {m_l^2}{2M_W^2 \cos^2\beta}|U_{i4}|^2\right )({\tilde M}_{\chi_i^+}^2-M_{{\tilde \nu}_L}^2-m_l^2)\right.\nonumber\\
&&\left.- 2\sqrt{2} \frac {m_l^2}{M_W \cos\beta} {\cal R}e[V_{i1} U_{i4}^{\ast}]{\tilde M}_{\chi_i^+} \right ], \nonumber \\
D(l^+ {\tilde \nu}_R)\hspace*{-0.2cm}&&=\frac{g^2}{2}\left [ \left (|U_{i2}|^2-\frac {m_l^2}{2M_W^2 \cos^2\beta}|V_{i4}|^2\right )({\tilde M}_{\chi_i^+}^2-M_{{\tilde \nu}_R}^2-m_l^2) \right. \nonumber \\
&&\left. - 2\sqrt{2} \frac {m_l^2}{M_W \cos\beta} {\cal R}e[U_{i2}^{\ast} V_{i4}]{\tilde M}_{\chi_i^+} \right ]. 
\end{eqnarray}
 With these expressions for the density matrices, the asymmetry becomes:
 \begin{eqnarray}
{\cal A}_l^T= \frac {\displaystyle\int {\rm Sign} {\cal T}_l \Sigma_P^2 \Sigma_D^2 \delta^4\left ( p_{e^+}+p_{e^-}-\sum \limits_i p_{i} \right)\prod \limits_i \frac {d^3 p_{i}} {(2 \pi)^3 2p_i^0} } {\displaystyle\int  |T|^2 \delta^4\left ( p_{e^+}+p_{e^-}-\sum \limits_i p_{i} \right)\prod \limits_i \frac {d^3 p_{i}} {(2 \pi)^3 2p_i^0} }, 
\end{eqnarray}
where $|T|^2=4| \Delta(\chi^+_i)|^2  (PD+\sum \limits_{a=1} ^3\Sigma_P^a \Sigma_D^a  )$. 
Using the narrow width approximation for the chargino propagator:
\begin{equation}
\Delta(\chi^{+}_{i})=\frac{1}{(s_i-{\tilde M}^2_{\chi_{i}}+i{\tilde M}_{\chi_{i}}\Gamma_{\chi_{i}})} \simeq \frac{\pi}{{\tilde M}_{\chi_i} \Gamma_{\chi_i}} \delta \left (s_i-{\tilde M}_{\chi_i}^2 \right) ,
\end{equation}
 the total cross section appearing in the denominator is obtained as 
\begin{equation} 
 d\sigma_{PD}=\frac{2}{s}PD (2 \pi)^4 \delta^4\left ( p_{e^+}+p_{e^-}-\sum \limits_i p_{i} \right) \prod \limits_i \frac {d^3 p_{i}} {(2 \pi)^3 2p_i^0}\,, 
 \end{equation}
and is polarization-independent.

An interesting sign of LRSUSY would be given by the case in which at least one of charginos could decay to an anti-lepton and a right-handed sneutrino. This might be an unlikely scenario, since $M_{{\tilde \nu}_R}$ is expected to be large. The splitting between the left and right sneutrinos is proportional to the Dirac neutrino mass, and it is conceivable that this is in the MeV region. If so, the right handed neutrino could be kinematically accessible for the decay of a heavier chargino. To achieve this, we consider a different mass scenario than in the previous section. We take the right-handed gaugino mass $|M_R|=600$ GeV, larger than the left-handed one ($|M_L|=150$ GeV). Under these circumstances, with the notation $\mathrm{{\tilde M}}_{\chi^{\pm}_{1}}\le\mathrm{{\tilde M}}_{\chi^{\pm}_{2}}\le\mathrm{{\tilde M}}_{\chi^{\pm}_{3}}\le\mathrm{{\tilde M}}_{\chi^{\pm}_{4}}$,  the fourth chargino $\chi_4$ is mainly a right-handed gaugino and it is heavy enough to decay to an anti-lepton and a right-handed sneutrino (taken to have mass of  500 GeV). The third lightest state is now mainly Higgsino and we will consider $\chi_3^+\to l^+\tilde{\nu}_L$ in conjunction with its production process from $e^+e^-$ scattering. Since  $A_l$ vanishes for the production of same charginos, we will only consider $e^+e^-\to \chi_1^-\chi_3^+\to\chi_1^- l^+\tilde{\nu}_L$ and $e^+e^-\to \chi_1^-\chi_4^+\to\chi_1^- l^+\tilde{\nu}_R$ processes, even though some other channels (with $\chi_i^+ \chi_i^-$ chargino final states) are allowed for cross section consideration.
\begin{figure}[t]
\vspace{-0.05in}  
    \centerline{\hspace*{3.1cm} \epsfxsize 3.5in {\epsfbox{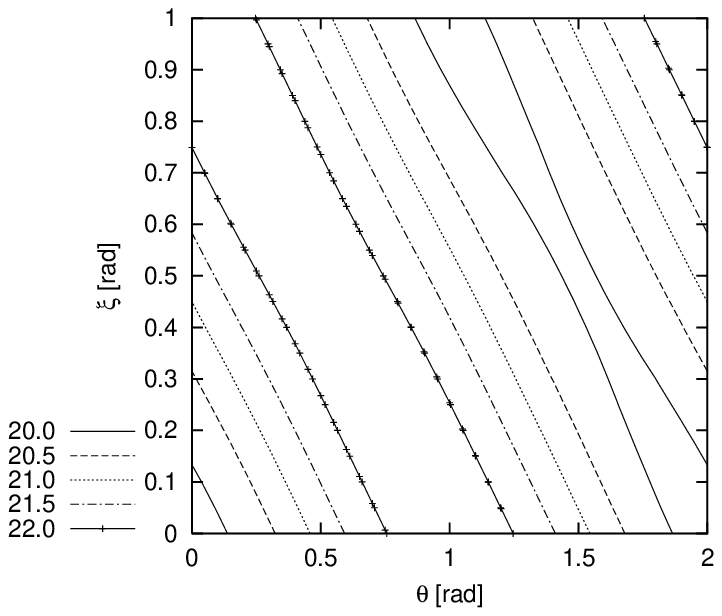}} \hspace{-1.5cm} \epsfxsize 3.5in {\epsfbox{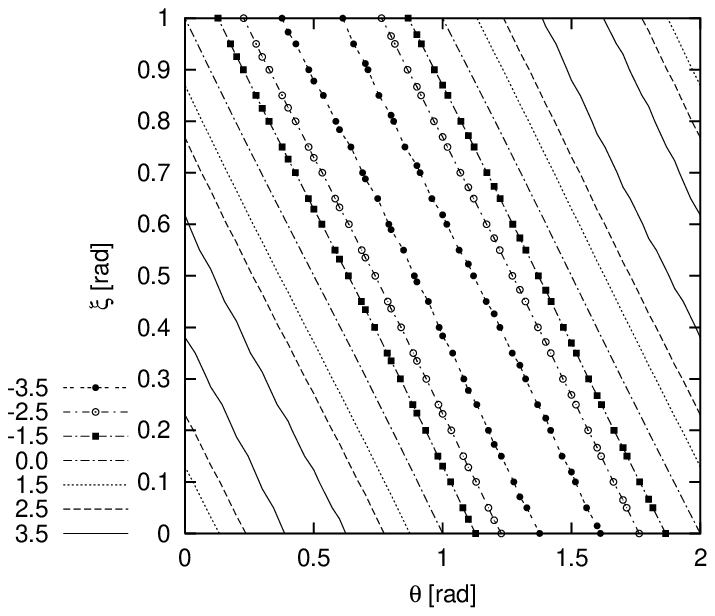}} }
\vskip -0.2in
    \caption{On the left, the total production-decay cross section $\sigma_{PD}=\sigma(e^+e^-\to\chi_3^+\chi_1^-)\times {\rm BR}(\chi_3^+\to l^+\tilde{\nu}_{L})$ ($l=e$ or $\mu$), in fb, in the CP-odd phases $(\theta,\xi)$-plane at $\sqrt{s}=1$ TeV, $M_R=600$ GeV, and $M_{\tilde{\nu}_{L}}=150$ GeV with a polarized electron-positron beam $(P_{e^-},P_{e^+})=(-0.8,0.6)\%$.  On the right, the T-odd asymmetry $A_l$ is depicted for the same process in the same plane and with the same parameter set. The asymmetry is given in percentage.}
    \label{fig:SigAToddL}
\end{figure}
\begin{figure}[htb]
\vspace{-0.05in}  
    \centerline{\hspace*{3.1cm} \epsfxsize 3.5in {\epsfbox{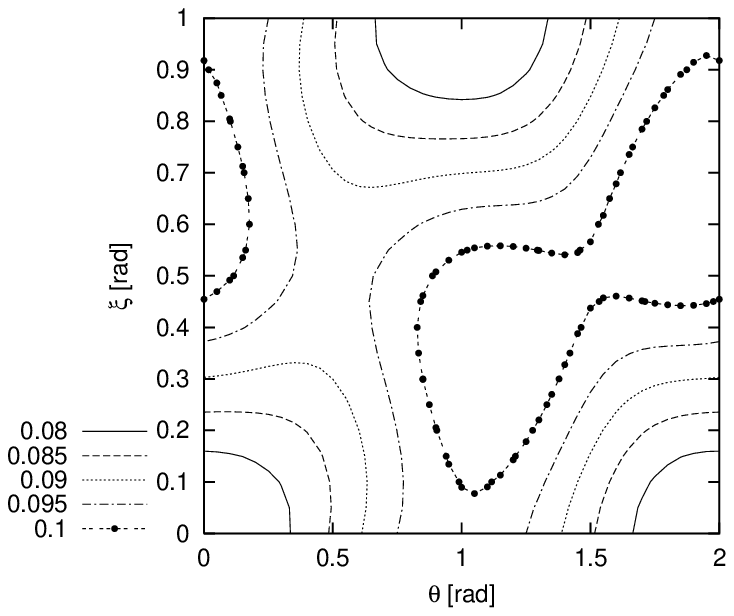}} \hspace{-1.5cm} \epsfxsize 3.5in {\epsfbox{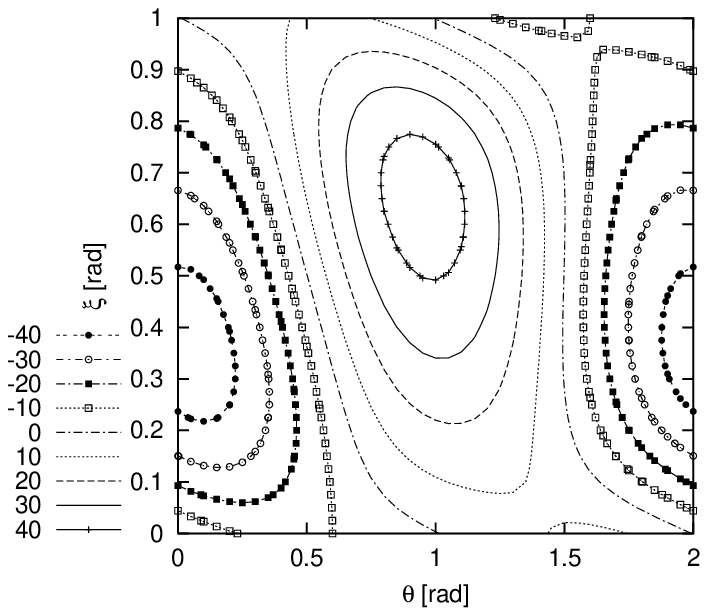}} }
\vskip -0.2in
    \caption{On the left, the total production-decay cross section $\sigma_{PD}=\sigma(e^+e^-\to\chi_4^+\chi_1^-)\times {\rm BR}(\chi_4^+\to l^+\tilde{\nu}_{R})$ ($l=e$ or $\mu$), in fb, in the CP-odd phases $(\theta,\xi)$-plane at $\sqrt{s}=1$ TeV, $M_R=600$ GeV, and $M_{\tilde{\nu}_{R}}=500$ GeV with a polarized electron-positron beam $(P_{e^-},P_{e^+})=(-0.8,0.6)\%$. On the right, the T-odd asymmetry $A_l$ is depicted for the same process in the same plane and with the same parameter set. The asymmetry is given in percentage.}
    \label{fig:SigAToddR}
\end{figure}

In Fig.~\ref{fig:SigAToddL}, we present, on the left, the total production-decay cross section $\sigma_{PD}=\sigma(e^+e^-\to\chi_3^+\chi_1^-)\times {\rm BR}(\chi_3^+\to l^+\tilde{\nu}_{L})$ ($l=e$ or $\mu$) at $\sqrt{s}=800$ GeV in the CP-odd phases $(\theta,\xi)$-plane with electron-positron beam polarization $(P_{e^-},P_{e^+})=(-0.8,0.6)\%$, and, on the right, the corresponding asymmetry $A_l$ for the same chain  with the same parameter set, defined as in the caption of the figure. The total cross section including the particular decay channel is $20$ fb without CP-odd phases and depends only weakly  on the phases. The maximum enhancement is around $10\%$. The cross section is approximately constant on the contours $\theta+\xi\sim {\rm const}$, and maximum enhancement occurs along such contours. There is a similar pattern in the asymmetry $A_l$ which varies in (-3.5\%,3.5\%) interval. CP-odd phases can give large asymmetry  in the entire $(\theta,\xi)$ range. $A_l$ takes values of  $3.5\%$ along $\theta+\xi\sim 0.5$ or $\theta+\xi\sim 2.5$, and $-3.5\%$ along $\theta+\xi\sim 1.5$.  These values for both the cross section and the asymmetry are pretty much consistent with the MSSM results, as expected. However, the process  
$e^+e^-\to \chi_1^-\chi_4^+\to\chi_1^- l^+\tilde{\nu}_R$ which is shown in Fig.~\ref{fig:SigAToddR}, is peculiar to the left right supersymmetric model and measuring the T-odd asymmetry associated with it could be used to distinguish it from MSSM. Even though the total cross section is comparatively small with respect to the previous case, the asymmetry  can reach $\pm 40\%$. Since the asymmetry is around $-6\%$ without the phases, the experimental availability of an asymmetry measurement would allow us to bound the CP-odd phases in the chargino sector of the model.

\section{Conclusion}
\label{sec:conclusion}

We have analysed the influence of the CP-violating phases in the chargino sector of the left-right supersymmetric model. Our emphasis has been mostly on the presence of a right-handed wino, and less so on additional higgsinos, which arise from (model-dependent) ways to break the symmetry. To achieve this, we decoupled the additional higgsinos by retaining only those originating from the bidoublet Higgs bosons, which are common to all left-right models. 

The added advantage of this method is that we are able to solve analytically for the chargino mass eigenvalues. From the most general set of phases in the mass matrix, two CP-odd phases arise (as opposed to just one in MSSM) and their influence is the strongest on the lightest chargino (mostly a mixture of left and right winos) mass, where effects of up to 20\% can be detected. The phases influence weakly the heaviest (corresponding to the MSSM higgsino) mass, and not much the other two states.

For the chargino production from $e^+e^-$ pairs,  a large CP-odd phase effect in the cross section is obtained for either same or different states production, at both $\sqrt{s}=500$ and $1000$ GeV. The variation of the cross section has distinct features: it is enhanced or depressed for a single region of combined parameter space for same charginos, while for different charginos the enhancement can occur for several parameter regions. The forward-backward asymmetry $A_{FB}$ (1.2\% at $\sqrt{s}=500$ GeV and 12\% $\sqrt{s}=1000$ GeV without CP phases) is a very sensitive probe of CP-odd angles, being enhanced to 50\% for $\sqrt{s}=500$ GeV for $\chi_1^+ \chi_1^-$ production. This is even more pronounced for $\chi_1^+ \chi_2^-$ where the maximum forward-backward asymmetry can be almost four times its value without CP phases. 

A distinguishing sign of a left-right gaugino sector would be the decay of charginos into left and right, if kinematic restrictions allow it, scalar neutrino. We study these decays seperately, in conjunction with the chargino production from a polarized electron-positron beam, and investigate the dependence of the cross section for chargino production and decay, as well as the CP asymmetry in this decay, $A_l$, as a function of the two CP-odd phases. Though for this decay to occur, the right-handed gaugino mass must be larger than the right sneutrino mass, and neither should decouple from the energy spectrum of the lower lying charginos, we find this possibility very interesting for detecting a signal for left-right supersymmetry. While the cross section for production and decay is enhanced, but not very sensitive to the CP-phases, the T-odd asymmetry is very sensitive to the phases. If the decay into the right-handed sneutrino is open, the enhancement of the asymmetry could be four times the one obtained in MSSM and would serve as a distinguishing signal for this model.

In summary, additional CP-odd phases arise in LRSUSY. Their effect is strongest in forward-backward asymmetry of the production, and in the CP-asymmetry in polarized production, followed by the decay into a neutrino and slepton. Both of these asymmetries show large enhancements with respect to their values in MSSM.
\begin{acknowledgments}
This work was funded by NSERC of Canada (SAP0105354).
\end{acknowledgments}


\end{document}